\documentclass[%
 reprint,
 superscriptaddress,
 amsmath,amssymb,
 aps,
 prfluids, 
 onecolumn,
 floatfix
]{revtex4-2}

\usepackage{graphicx}
\usepackage{siunitx}

\usepackage[usenames,dvipsnames]{xcolor}

\newcommand{\hypgeo}[2]{%
  \operatorname{%
    {\vphantom{\mathnormal{F}}}_{#1}%
    \kern-\scriptspace
    \mathnormal{F}_{#2}%
  }%
}

\usepackage[%
    linkcolor = blue, 
    citecolor = blue, 
    urlcolor = blue, 
    colorlinks = true
]{hyperref}

\begin{document}


\title{Diffusioosmotic flow in a soft microfluidic configuration induces fluid--structure instability}

\author{Nataly Maroundik}
\affiliation{Faculty of Mechanical Engineering, Technion -- Israel Institute of Technology, Haifa 3200003, Israel}
\author{Dotan Ilssar}
\affiliation{Department of Mechanical and Process Engineering, ETH Z{\"u}rich, Z{\"u}rich 8092, Switzerland}
\author{Evgeniy Boyko}
\email[Email address for correspondence: ]{evgboyko@technion.ac.il}
\affiliation{Faculty of Mechanical Engineering, Technion -- Israel Institute of Technology, Haifa 3200003, Israel}

\date{\today}

\begin{abstract}

Diffusioosmotic flow arises in microfluidic configurations due to solute concentration gradients. In soft microfluidic channels, internal pressure gradients generated by diffusioosmotic flow to conserve mass result in elastic deformation of the channel walls, triggering fluid--structure interaction. In this work, we analyze the fluid--structure interaction between diffusioosmotic flow of an electrolyte solution and a deformable microfluidic channel. We provide insight into the physical behavior of the system by developing a reduced-order model, in which a viscous film is confined between a rigid bottom surface and an elastic top substrate, represented as a rigid plate connected to a linear spring. Considering a slender configuration and applying the lubrication approximation, we derive a set of two-way coupled governing equations describing the evolution of the fluidic film thickness and the solute concentration. Our theoretical predictions show that above a certain concentration gradient threshold, negative pressures induced by diffusioosmotic flow give rise to fluid--structure instability, causing the elastic top substrate to collapse onto the bottom surface. We employ theoretical analysis to elucidate the underlying physical mechanisms for the onset of fluid--structure instability by performing a linear stability analysis of the system and identifying three distinct dynamic regimes. We validate our theoretical results with finite-element simulations and find excellent agreement. The understanding of this instability is of fundamental importance for improving the control and design of microfluidic systems driven by diffusioosmotic flow and containing soft elements.
\end{abstract}

\maketitle 

\section{Introduction}

Diffusioosmotic flow or diffusioosmosis is the spontaneous movement of fluid adjacent to a surface, driven by a solute concentration gradient~\cite{prieve1984motion,anderson1989colloid,shim2022diffusiophoresis,ault2024physicochemical}. In electrolyte solutions, diffusioosmotic flow arises from two contributions. The first contribution is chemiosmosis, which is driven by osmotic pressure gradients along the surface~\cite{anderson1989colloid,ault2024physicochemical}. The second contribution is electroosmosis, which can be induced by an interaction of a spontaneously generated electric field with the net charge in the electric double layer, due to unequal ion diffusivities~\cite{anderson1989colloid,ault2024physicochemical}. \citet{prieve1984motion}~developed a classical model for diffusioosmotic flow of electrolyte solutions and provided an analytical expression for the diffusioosmotic slip velocity at a solid surface in the thin-double-layer limit. Recently,~\citet{shi2025generalized} presented a generalized, conceptually unified theory of linear osmotic slip velocity over a solid surface, building on Derjaguin’s original approach~\cite{derjaguin1961diffusiophoresis,derjaguin1993diffusiophoresis} and reproducing known formulas for linear electroosmosis, chemiosmosis, and diffusioosmosis.

With the advent of microfabrication technology in the 2000s, diffusioosmotic flow has become a common driving mechanism for fluid manipulation in microfluidic and lab-on-a-chip devices~\cite{lee2014osmotic,shim2022diffusiophoresis,ault2024physicochemical,liu2025diffusioosmotic}, alongside pressure-driven and electroosmotic flows~\cite{stone2004engineering,squires2005microfluidics,paratore2022reconfigurable}. For example,~\citet{ault2019characterization} presented a microfluidic approach that leverages diffusioosmotic flow to measure the zeta potentials of electrolyte solutions. Recently,~\citet{teng2023diffusioosmotic} studied solute dispersion driven solely by diffusioosmotic flow. A closely related phenomenon to diffusioosmosis is diffusiophoresis, which refers to the spontaneous phoretic motion of suspended particles in a solute concentration gradient due to slip flow at the surface~\cite{derjaguin1961diffusiophoresis,derjaguin1993diffusiophoresis,prieve1984motion,anderson1989colloid}. The physical origin of diffusiophoresis is analogous to that of diffusioosmotic flow and results from chemiphoretic and electrophoretic contributions in electrolyte solutions~\cite{prieve1984motion,anderson1989colloid}. Since its discovery by~\citet{derjaguin1961diffusiophoresis}, diffusiophoresis has been widely studied as a driving mechanism for the migration and manipulation of particles~\cite{ault2017diffusiophoresis,gupta2020diffusiophoresis,singh2020reversible,singh2022enhanced,shim2022diffusiophoresis,migacz2024enhanced,ault2024physicochemical}. In particular, understanding the coupled effects of diffusioosmosis and diffusiophoresis on particle dynamics in various microfluidic configurations, including dead-end channels and T-junctions, has received much attention in the fluid mechanics and soft matter communities through analytical modeling, numerical simulations, and experiments~\cite{kar2015enhanced,shin2017accumulation,shin2016size,ault2018diffusiophoresis,ha2019dynamic,alessio2021diffusiophoresis,alessio2022diffusioosmosis,akdeniz2023diffusiophoresis,liu2025diffusioosmotic}. We refer the reader to the recent review articles by~\citet{shim2022diffusiophoresis} and~\citet{ault2024physicochemical} and the references therein, for a further discussion on diffusioosmotic flow and diffusiophoresis.

Microfluidic configurations are often fabricated from soft materials such as poly(dimethylsiloxane) (PDMS)~\cite{whitesides2006origins}, and as a result, viscous flows through such configurations may induce elastic deformation of the channel cross-section~\cite{christov2021soft}. 
Elucidating the fluid--structure interaction between internal fluid flow and the elastic boundaries of deformable channels plays a central role in understanding hydrodynamic behavior across a wide range of applications, including  wearable diagnostics~\citep{kashaninejad2023microfluidic}, biomedical applications~\citep{grotberg2004biofluid,bhatia2014microfluidic}, control of viscous fingering~\citep{al2013controlling,fontana2024peeling},
fabrication of microfluidic devices~\citep{gervais2006flow,dendukuri2007stop,hardy2009deformation,christov2018flow,christov2021soft}, soft robotics~\citep{shepherd2011multigait,elbaz2014dynamics,rus2015design,polygerinos2017soft,MEG17,salem2020leveraging}, and flow control in plants’ vasculature~\citep{park2021fluid}.
For instance,~\citet{elbaz2014dynamics} explored the transient fluid--structure interactions of pressure-driven flows in axisymmetric geometries, demonstrating their relevance to soft robotics applications. \citet{christov2018flow} employed lubrication theory along with the Kirchhoff--Love bending model to derive the relationship between flow rate and pressure drop in a deformable microchannel, showing that the flow rate--pressure drop relation exhibits a sublinear behavior due to wall compliance.
Recently,~\citet{boyko2022flow} demonstrated that the reciprocal theorems for Stokes flow and linear elasticity can be leveraged to derive a closed-form expression for the flow rate--pressure drop relation in deformable channels, thereby bypassing the need to solve the full fluid--structure interaction problem.~\citet{martinez2020start} extended the steady-state analysis of~\citet{christov2018flow} to the start-up flow in deformable microchannels by incorporating fluid and solid inertia into the lubrication and Kirchhoff--Love bending equations,
while~\citet{guyard2022elastohydrodynamic} studied the elastohydrodynamic relaxation behavior of such configurations.
Furthermore,~\citet{biviano2022smoothing} explored how the compliance of deformable microchannels can be harnessed to attenuate oscillations in peristaltic pump flow using bioinspired passive components. 

In tandem with studies on pressure-driven flows through deformable microchannels~\citep{christov2021soft}, there has also been growing interest in exploring the role of alternative actuation mechanisms, such as electroosmotic flow, in driving fluid--structure interactions at the microscale~\citep{mukherjee2013relaxation,de2016numerical,rubin2017elastic,boyko2019elastohydrodynamics}. In particular,~\citet{rubin2017elastic} and~\citet{boyko2019elastohydrodynamics} examined the use of non-uniform electroosmotic flow as a driving mechanism to generate desired dynamic deformations in a lubricated elastic sheet by inducing internal pressure gradients within the fluid. Assuming small deformations and strong pre-stretching of the elastic sheet,~\citet{boyko2019elastohydrodynamics} analyzed the linear and weakly nonlinear viscous--elastic interaction driven by non-uniform electroosmotic flow, which showed stable behavior. 
For a further discussion on low-Reynolds-number fluid--structure interactions, we refer the reader to the overviews given
recently by~\citet{christov2021soft} and~\citet{rallabandi2024fluid}.

It should be noted that in all the studies mentioned above, the fluid--structure interactions exhibited stable behavior. However, the study of low-Reynolds-number fluid--structure interactions is not limited to stable scenarios. Indeed, various fluid--structure instabilities of the fluid--elastic interface have been reported, highlighting the rich dynamics beyond stable behavior~\citep{heil2011fluid,juel2018instabilities,christov2021soft}. These include viscous flows in collapsible tubes~\citep{heil1997stokes}, 
snap-through instabilities arising from the interaction between a buckled elastic arch and viscous flow~\citep{gomez2017passive}, wrinkling of lubricated elastic sheets under compression~\citep{kodio2017lubricated}, underactuated control of multistable elastic membranes using viscous fluids~\citep{peretz2020underactuated}, and flow-induced choking in compliant Hele-Shaw cells~\citep{box2020flow}. Of particular interest are the studies by~\citet{boyko2020nonuniform} and~\citet{boyko2020interfacial} on fluid--structure interactions driven by non-uniform electroosmotic flow. They showed that above a certain critical value of the imposed electric field, negative pressures formed by non-uniform electroosmotic flow can give rise to the instability of the liquid--elastic interface. Even though diffusioosmotic flow is a common driving mechanism at the microscale, to the best of the authors' knowledge, neither the fluid--structure interaction nor the fluid--structure instability that may be induced by diffusioosmotic flow in deformable channels has been studied to date, thereby motivating the present study.

In this theoretical work, we analyze the low-Reynolds-number fluid--structure interaction between diffusioosmotic flow and a deformable microchannel. 
We provide insight into the physical behavior of the system by developing a simplified model, in which a viscous film is lubricated between a rigid bottom surface and an elastic top substrate, modeled as a rigid plate connected to a linear spring. Diffusioosmotic flow, driven by a solute concentration difference at the edges, induces a fluidic pressure exerted on the plate, leading to the fluid--structure interaction. 
Since the pressure generated by diffusioosmotic flow scales inversely with the fluidic film thickness, as we show, exceeding a certain solute concentration difference between the edges can result in sufficiently negative pressures to trigger fluid–structure instability, leading to the collapse of the elastic top substrate onto the bottom surface. 

The paper is organized as follows.  In Sec.~\ref{Sec: Problem_formulation}, we present the problem formulation and governing equations for the fluid dynamics and the solute transport. In Sec.~\ref{Sec: Reduced_order_model}, we present a reduced-order model that describes fluid--structure interaction driven by diffusioosmotic
flow. We derive a pair of two-way coupled equations governing the evolution of the fluidic film thickness and the solute concentration. We provide their scaling and introduce key non-dimensional parameters. 
In Sec.~\ref{Sec: Results}, we present our theoretical results for the fluid--structure interaction dynamics and demonstrate the onset of fluid--structure instability driven by diffusioosmotic flow. We elucidate the underlying physical mechanisms that lead to the onset of fluid--structure instability and identify three distinct dynamic regimes. We further present a linear stability analysis of the system and rationalize the dynamic predictions of the theoretical model. We validate the results of our theoretical model by performing two-dimensional finite-element numerical simulations, finding excellent
agreement. Finally, we conclude with a discussion of the results in Sec.~\ref{CR}.

\begin{figure}[t]
   \centering
    \includegraphics[scale=1.1]{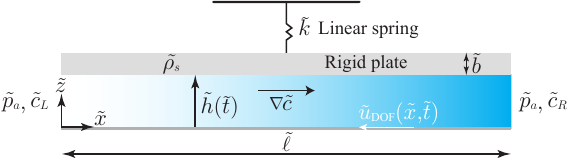}
    \caption{Schematic illustration of the modeled configuration, showing the coordinate system and relevant physical and geometric parameters. A thin viscous film of initial thickness $\tilde{h}_i$ is confined between a rigid bottom surface and a rigid top plate, connected to a linear spring of stiffness $\tilde{k}$. A solute concentration difference imposed at the edges generates a local concentration gradient, inducing a diffusioosmotic slip velocity $\tilde{\boldsymbol{u}}_{\rm DOF}(\tilde{x}, \tilde{t})$ along the bottom surface. This slip velocity drives a diffusioosmotic flow, which generates an internal fluidic pressure distribution acting on the top plate, resulting in fluid--structure interaction and the time evolution of the film thickness $\tilde{h}(\tilde{t})$.}
    \label{F1}
\end{figure}

\section{Problem formulation and governing equations}\label{Sec: Problem_formulation}

We study the fluid--structure interaction and instability of a viscous fluid with density $\tilde{\rho}$, viscosity $\tilde{\mu}$, and permittivity $\tilde{\varepsilon}$, driven by diffusioosmotic flow and confined between a rigid bottom surface and a rigid top plate. The top plate has density $\tilde{\rho}_s$, thickness $\tilde{b}$, and length $\tilde{\ell}$, and is connected to a linear spring with stiffness $\tilde{k}$. The initial film thickness, representing the initial gap between the flat surfaces, is given by $\tilde{h}_i$.
We consider a planar two-dimensional configuration and employ a Cartesian coordinate system $(\tilde{x},\tilde{z})$, whose $\tilde{x}$ axis lies at the bottom flat surface and $\tilde{z}$ is perpendicular thereto, as shown in Fig.~\ref{F1}. 

We assume that, initially, the solute concentration $\tilde{c}$ is uniform, such that $\tilde{c}(\tilde{x}, \tilde{z}, \tilde{t} = 0) = \tilde{c}_L$, where $\tilde{t}$ denotes time.
We induce a concentration gradient by imposing a solute concentration difference between the left and right edges, $\tilde{c}(\tilde{x}=0, \tilde{z}, \tilde{t} ) = \tilde{c}_L$ and $\tilde{c}(\tilde{x}=\tilde{\ell}, \tilde{z}, \tilde{t} ) = \tilde{c}_R$, which generates a diffusioosmotic slip velocity at the bottom surface, $\tilde{\boldsymbol{u}}_{\rm DOF}(\tilde{x},\tilde{t})=-\tilde{\Gamma} \boldsymbol{\nabla}_{\Vert}\ln \tilde{c}$~\cite{prieve1984motion,anderson1989colloid,ault2024physicochemical}, where $\tilde{\Gamma}$ is the diffusioosmotic mobility and $\boldsymbol{\nabla}_{\Vert}=(\partial/\partial \tilde{x})\hat{\boldsymbol{x}}$ is the gradient is parallel to the surface. This slip velocity drives a diffusioosmotic flow with pressure distribution $\tilde{p}$ and velocity $\tilde{\boldsymbol{u}} = (\tilde{u}, \tilde{w})$, which in turn causes the time evolution of the gap $\tilde{h}(\tilde{t})$ and leads to solute advection in addition to diffusion. The characteristic velocities in the $\hat{\boldsymbol{x}}$ and $\hat{\boldsymbol{z}}$ directions are $\tilde{u}_c$ and $\tilde{w}_c$, respectively, while the characteristic pressure and time are denoted by $\tilde{p}_c$ and $\tilde{t}_c$.
For simplicity, we assume that the influence of gravity is
negligible.

The time evolution of the gap $\tilde{h}(\tilde{t})$ is governed by the coupled dynamics of the fluid motion and solute transport within the channel.
In this work, we focus on slender configurations, where $\tilde{h}(\tilde{t})\ll \tilde{\ell}$, and assume negligible fluidic inertia, represented by small Womersley and reduced Reynolds numbers,
\begin{equation}
\epsilon=\frac{\tilde{h}_{i}}{\tilde{\ell}}\ll1,\qquad Wo=\frac{\tilde{\rho}\tilde{h}_{i}^2}{\tilde{\mu} \tilde{t}_c}\ll1,\qquad\epsilon Re=\epsilon\frac{\tilde{\rho}\tilde{u}_{c}\tilde{h}_{i}}{\tilde{\mu}}\ll1,\label{Lubrication assumptions}
\end{equation}
corresponding to the assumptions of the classical lubrication approximation~\cite{leal2007advanced}.
Therefore, the fluid motion is governed by the lubrication equations~\cite{leal2007advanced}
\begin{subequations} \label{Lubrication_equations}
\begin{align}
\tag{\theparentequation a} \label{Lubrication_a}
\frac{\partial \tilde{u}}{\partial \tilde{x}} + \frac{\partial \tilde{w}}{\partial \tilde{z}} &= 0, \\
\tag{\theparentequation b} \label{Lubrication_b}
\frac{\partial \tilde{p}}{\partial \tilde{x}} &= \tilde{\mu} \frac{\partial^2 \tilde{u}}{\partial \tilde{z}^2}, \\
\tag{\theparentequation c} \label{Lubrication_c}
\frac{\partial \tilde{p}}{\partial \tilde{z}}& = 0,
\end{align}
\end{subequations}
and the solute transport is governed by the advection--diffusion equation
\begin{equation}
    \frac{\partial \tilde{c}}{\partial \tilde {t}} + \tilde{u}\frac{\partial \tilde{c}}{\partial \tilde{x}}+\tilde{w}\frac{\partial \tilde{c}}{\partial \tilde{z}} = \tilde{D}\left(\frac{\partial^{2} \tilde{c}}{\partial \tilde{x}^2} + \frac{\partial^{2} \tilde{c}}{\partial \tilde{z}^2}\right), \label{adv-diff}
\end{equation}
where $\tilde{D}$ is the solute diffusivity. From  Eq.~(\ref{Lubrication_c}), it follows that $\tilde{p}=\tilde{p}(\tilde{x},\tilde{t})$, i.e., the pressure is uniform across the gap and varies only along the axial direction and in time.

These governing equations are subject to the following boundary conditions on the fluid and solute
\begin{subequations}
\begin{align}
    \text{Diffusioosmotic slip at the bottom surface:} \quad 
    & \tilde{u}(\tilde{x},\tilde{z}=0,\tilde{t}) = \tilde{u}_{\rm DOF}(\tilde{x},\tilde{t}) 
    = -\tilde{\Gamma} \left. \frac{\partial \ln \tilde{c}}{\partial \tilde{x}} \right|_{\tilde{z}=0}, \label{BC_DOS} \\
    \text{No fluid penetration at the bottom surface:} \quad 
    & \tilde{w}(\tilde{x},\tilde{z}=0,\tilde{t}) = 0, \label{BC_no_penetration_bottom} \\
    \text{No slip at the top plate:} \quad 
    & \tilde{u}(\tilde{x},\tilde{z}=\tilde{h}(\tilde{t}),\tilde{t}) = 0, \label{BC_no_slip_plate} \\
    \text{Kinematic condition at the top plate:} \quad 
    & \tilde{w}(\tilde{x},\tilde{z}=\tilde{h}(\tilde{t}),\tilde{t}) = \frac{d \tilde{h}}{d \tilde{t}}, \label{BC_kinamatic_plate} \\
    \text{Fixed pressure at the edges:} \quad 
    & \tilde{p}(\tilde{x}=0, \tilde{z}, \tilde{t}) = \tilde{p}(\tilde{x}=\tilde{\ell},\tilde{z}, \tilde{t}) = 0, \label{BC_pressure} \\
    \text{Fixed concentration at the edges:} \quad 
    & \tilde{c}(\tilde{x}=0, \tilde{z}, \tilde{t}) = \tilde{c}_{L}, \quad 
      \tilde{c}(\tilde{x}=\tilde{\ell}, \tilde{z}, \tilde{t}) = \tilde{c}_R, \label{D_concentartion_bc} \\
    \text{No-flux conditions at the bottom and top surfaces:} \quad 
    & \left.\frac{\partial \tilde{c}}{\partial \tilde{z}}\right|_{\tilde{z}=0} = 0, \quad 
      \tilde{D} \left.\frac{\partial \tilde{c}}{\partial \tilde{z}}\right|_{\tilde{z}=\tilde{h}(\tilde{t})} 
      = \frac{d\tilde{h}}{d\tilde{t}} \tilde{c}|_{\tilde{z}=\tilde{h}(\tilde{t})}. \label{D_bc_conc}
\end{align}
\end{subequations}
The slip boundary condition (\ref{BC_DOS}) represents the diffusioosmotic slip velocity, which serves as the driving mechanism for the fluid--structure interaction in our study. This diffusioosmotic flow arises from solute concentration gradients introduced into the system through the boundary conditions~(\ref{D_concentartion_bc}), where $\tilde{c}_R$ is the prescribed solute concentration at the right edge. The last boundary condition (\ref{D_bc_conc}) represents the no-flux conditions at the bottom and top surfaces, i.e., $\tilde{\boldsymbol{j}} \cdot\hat{\boldsymbol{n}}=0$ at $\tilde{z}=0, \tilde{h}(\tilde{t})$, where $ \tilde{\boldsymbol{j}} = \tilde{c}\tilde{\boldsymbol{u}}-\tilde{D}\boldsymbol{\nabla}\tilde{c}$ is the molar flux. Noting that $\hat{\boldsymbol{n}}= \pm\hat{\boldsymbol{z}}$, the latter condition leads to $\tilde{\boldsymbol{j}} \cdot\hat{\boldsymbol{n}} = \tilde{w}\tilde{c}-\tilde{D}\partial \tilde{c}/\partial \tilde{z}=0$ at $\tilde{z}=0, \tilde{h}(\tilde{t})$.
Thus, using Eqs.~(\ref{BC_no_penetration_bottom}) and (\ref{BC_kinamatic_plate}), the no-flux boundary conditions take the form given in Eq.~(\ref{D_bc_conc}).

As initial conditions, we assume a uniform solute concentration throughout the channel,
\begin{equation}
     \tilde{c}(\tilde{x},\tilde{z},\tilde{t}=0) = \tilde{c}_
     L.\label{IC_conc2}
\end{equation}
and an initial gap height,
\begin{equation}
    \tilde{h}(\tilde{t} = 0) = \tilde{h}_{i}.
\end{equation}

In this work, we assume a constant zeta potential and diffusioosmotic mobility coefficient $\tilde{\Gamma}$, which is a reasonable first approximation under many scenarios, as discussed by recent previous studies~\cite{migacz2022diffusiophoresis,lee2023role,akdeniz2023diffusiophoresis}.
This assumption is widely used in the literature~(see, e.g.,~\cite{ault2018diffusiophoresis,ault2019characterization,teng2023diffusioosmotic,liu2025diffusioosmotic}), as it enables the derivation of closed-form analytical results and is typically valid when the ratio of solute concentrations in the system is not too large (i.e., $\tilde{c}_R/\tilde{c}_L$ is not too large).

In addition, we neglect any dispersion because the effects
of diffusioosmotic flow on the dispersion are $O(\epsilon^2)$. Indeed, recent studies by~\citet{alessio2022diffusioosmosis}~and~\citet{teng2023diffusioosmotic} have demonstrated that the effective diffusivity $\tilde{D}_{\rm eff}$ induced by diffusioosmotic flow in a two-dimensional channel is given by
\begin{equation}
   \tilde{D}_{\rm eff} =  \tilde{D} \left[1+\left(\frac{\tilde{\Gamma}}{\tilde{D}}\right)^2 \frac{\epsilon^2}{210}\left(\frac{\partial \ln \tilde{c}}{\partial \tilde{x}}\right)^2+O(\epsilon^4)\right], \label{D_eff}
\end{equation}
and therefore, the effective diffusivity can be approximated by the solute diffusivity to the leading order in $\epsilon$.

\section{Reduced-order model for fluid--structure interaction driven by diffusioosmotic flow }\label{Sec: Reduced_order_model}

In this section, we develop a reduced-order model to describe fluid--structure interaction driven by diffusioosmotic flow, and derive the governing equations for the coupled evolution of solute concentration and gap height.

\subsection{Evolution equation for the film thickness}\label{Subsec: Evolution_film_thickness}

Applying the lubrication approximation and using Eqs.~(\ref{Lubrication_equations}) and (\ref{BC_DOS})--(\ref{BC_kinamatic_plate}), the evolution of the film thickness $\tilde{h}(\tilde{t})$ is related to the fluidic pressure $\tilde{p}(\tilde{x},\tilde{t})$ by the Reynolds equation (see p. 313 in~\cite{leal2007advanced})
\begin{equation}
    \frac{d \tilde{h}(\tilde{t})}{d \tilde{t}} -\frac{\tilde{h}(\tilde{t})^3}{12\tilde{\mu}} \frac{\partial^2 \tilde{p}(\tilde{x},\tilde{t})}{\partial \tilde{x}^2} =-\frac{\tilde{h}(\tilde{t})}{2} \frac{\partial \tilde{u}_{\rm DOF}(\tilde{x},\tilde{t})}{\partial \tilde{x}}. \label{D Reynolds with h=h(t)}
\end{equation}
The term on the right-hand side of Eq.~(\ref{D Reynolds with h=h(t)}) represents spatial variations in diffusioosmotic flux $\tfrac{1}{2}\tilde{h}\tilde{u}_{\rm DOF}$, which drives the fluid--structure interaction.
Solving the Reynolds equation~(\ref{D Reynolds with h=h(t)}) subject to Eq.~(\ref{BC_pressure}) yields the fluidic pressure $\tilde{p}(\tilde{x},\tilde{t})$,
\begin{equation}
    \tilde{p}(\tilde{x},\tilde{t})=\frac{6\tilde{\mu}}{\tilde{h}^3}\frac{d\tilde{h}}{d\tilde{t}}\tilde{x}(\tilde{x}-\tilde{\ell})+\frac{6\tilde{\mu}}{\tilde{h}^2}\left(\tilde{A}(\tilde{x},\tilde{t})-\frac{\tilde{x}}{\tilde{\ell}}\tilde{A}(\tilde{\ell},\tilde{t})\right),\label{D_pressure}
\end{equation}
where $\tilde{A}(\tilde{x},\tilde{t})$ is defined as 
\begin{equation}
    \tilde{A}(\tilde{x},\tilde{t})=\int_{0}^{\tilde{x}} \tilde{u}_{\rm DOF}(\tilde{x},\tilde{t})d\tilde{x}=-\tilde{\Gamma} \ln\left(\frac{\tilde{c}(\tilde{x},\tilde{z}=0,\tilde{t})}{ \tilde{c}_{L}}\right).\label{D_A}
\end{equation}

Integrating Eq.~(\ref{Lubrication_b}) twice with respect to $\tilde{z}$ and applying the boundary conditions (\ref{BC_DOS}) and (\ref{BC_no_slip_plate}) at the bottom and top surfaces, we obtain the expression for axial velocity $\tilde{u}(\tilde{x},\tilde{z},\tilde{t})$. Substituting this result into the definition of the  flow rate $\tilde{q}(\tilde{x},\tilde{t})=\int_{0}^{\tilde{h}(\tilde{t})}\tilde{u}d\tilde{z}$ and using Eqs.~(\ref{D_pressure})--(\ref{D_A}), the flow rate can be expressed as
\begin{equation}
    \tilde{q}(\tilde{x},\tilde{t}) =-\frac{\tilde{h}^3}{12\tilde{\mu}} \frac{\partial \tilde{p}}{\partial \tilde{x}} +\frac{\tilde{h}}{2} \tilde{u}_{\rm DOF}=\tilde{\ell}\frac{d\tilde{h}}{d\tilde{t}}\left(\frac{1}{2}-\frac{\tilde{x}}{\tilde{\ell}}\right)+ \frac{\tilde{h}}{2{\tilde{\ell}}}\tilde{A}(\tilde{\ell},\tilde{t})=\tilde{\ell}\frac{d\tilde{h}}{d\tilde{t}}\left(\frac{1}{2}-\frac{\tilde{x}}{\tilde{\ell}}\right)- \frac{\tilde{h}}{2{\tilde{\ell}}}\tilde{\Gamma}\ln\left(\frac{\tilde{c}_R}{\tilde{c}_L}\right).\label{D_flow_rate}
\end{equation}
We note that, in the case of $\tilde{h}=\text{const.}$, Eq.~(\ref{D_flow_rate}) is consistent with the expression for the diffusioosmotic flow rate in a channel, as derived by~\citet{lee2014osmotic}.
Integrating Eq.~(\ref{D_pressure}) with respect to $\tilde{x}$ from 0 to $\tilde{\ell}$ provides the expression for the fluidic force $\tilde{F}_f(\tilde{t})=\int_{0}^{\tilde{\ell}} \tilde{p} d\tilde{x}$,
\begin{equation}
    \tilde{F}_{\it f}(\tilde{t}) = -\frac{\tilde{\mu}\tilde{\ell}^3}{\tilde{h}(\tilde{t})^3}\frac{d\tilde{h}(\tilde{t})}{d\tilde{t}} + \frac{6\tilde{\mu}\tilde{B}(\tilde{t})}{\tilde{h}(\tilde{t})^2},\label{D_Ff}
\end{equation}
where $\tilde{B}(\tilde{t})$ is defined as
\begin{equation}
    \tilde{B}(\tilde{t}) = \int_{0}^{\tilde{\ell}} \left[\tilde{A}(\tilde{x},\tilde{t})-\frac{\tilde{x}}{\tilde{\ell}}\tilde{A}(\tilde{\ell},\tilde{t})\right]d\tilde{x} = \int_{0}^{\tilde{\ell}}\int_{0}^{\tilde{x}'} \tilde{u}_{\rm DOF}(\tilde{x}',\tilde{t})d\tilde{x}'d\tilde{x} - \frac{\tilde{\ell}}{2}\int_{0}^{\tilde{\ell}} \tilde{u}_{\rm DOF}(\tilde{x},\tilde{t})d\tilde{x}.\label{D_B(x,t)}
\end{equation}
The first term on the right-hand side of Eq.~(\ref{D_Ff}) represents the viscous resistance and the second term represents the diffusioosmotic force, which can be either attractive or repulsive depending on the sign of $\tilde{B}$.

Substituting the expression for the diffusioosmotic slip velocity (\ref{BC_DOS}) into Eq.~(\ref{D_B(x,t)}) and using the boundary conditions (\ref{D_concentartion_bc}), $\tilde{B}(\tilde{t})$ takes the form
\begin{equation}
    \tilde{B}(\tilde{t}) = -\tilde{\Gamma}\int_{0}^{\tilde{\ell}}\ln\left(\frac{\tilde{c}(\tilde{x},\tilde{z}=0,\tilde{t})}{ \tilde{c}_{L}}\right)d\tilde{x}+\frac{\tilde{\ell}}{2}\tilde{\Gamma}\ln\left(\frac{\tilde{c}_R}{\tilde{c}_{L}}\right).\label{B_with_ln(c)}
\end{equation}
In addition to the fluidic force, an elastic force $\tilde{F}_e$ acts on the top plate, which we model as a linear restoring force proportional to the variation in the film thickness
\begin{equation}
    \tilde{F}_{\it e}=\tilde{k}(\tilde{h}_i-\tilde{h}(\tilde{t})).\label{D_Fe}
\end{equation}
Considering the force balance acting on the plate and accounting for the fluidic force Eq.~(\ref{D_Ff}) and the elastic force Eq.~(\ref{D_Fe}), we obtain the governing equation for the plate motion
\begin{equation}
  \tilde{\rho}_s\tilde{b}\tilde{\ell} \frac{d^2 \tilde{h}}{d\tilde{t}^2}=\tilde{F}_{\it f}+\tilde{F}_{\it e}.\label{force_eq}
\end{equation}
In this study, we focus on the viscous--elastic time scale $(\tilde{\ell}/\tilde{h}_{i})^{3}(\tilde{\mu}/\tilde{k})$~(see, e.g., ~\cite{elbaz2014dynamics,boyko2020nonuniform}), which is typically much longer than the characteristic elastic inertial time scale and therefore we neglect solid inertia and any vibrations it may induce.
Neglecting the solid inertia, $\tilde{\rho}_s \tilde{b}\tilde{\ell}/(\tilde{k}\tilde{t}_c^2)\ll 1$, and substituting the expressions for $\tilde{F}_f$ and $\tilde{F}_e$, given in Eqs.~(\ref{D_Ff}) and (\ref{D_Fe}), into Eq.~(\ref{force_eq}), yields the evolution equation for $\tilde{h}(\tilde{t})$
\begin{equation}
\underbrace{-\frac{\tilde{\mu}\tilde{\ell}^3}{\tilde{h}(\tilde{t})^3}\frac{d\tilde{h}(\tilde{t})}{d\tilde{t}}}_{\text{Viscous resistance}} 
+ \underbrace{\frac{6\tilde{\mu}\tilde{B}(\tilde{t})}{\tilde{h}(\tilde{t})^2}}_{\substack{\text{Diffusioosmotic}\\\text{forcing}}} 
+ \underbrace{\tilde{k}(\tilde{h}_i-\tilde{h}(\tilde{t}))}_{\text{Elastic forcing}} 
= 0. \label{force_eq_sub}
\end{equation}
Equation (\ref{force_eq_sub}) is a nonlinear governing equation for $\tilde{h}(\tilde{t})$, which accounts for viscous, diffusioosmotic, and elasticity effects, and describes the gap evolution driven by diffusioosmotic flow.

We note that similar nonlinear evolution equations are encountered in various instability problems, such as electrostatic MEMS actuators~\cite{rebeiz2004rf}, elastocapillary coalescence~\cite{singh2014fluid}, and fluid--structure instability driven by non-uniform electro-osmotic flow~\cite{boyko2020nonuniform}. 
For non-uniform electro-osmotic flow, the fluid--structure interaction is described by a single evolution equation for $\tilde{h}(\tilde{t})$~\cite{boyko2020nonuniform}.
However, for diffusioosmotic flow, the evolution of the gap depends on a time-varying function~$\tilde{B}(\tilde{t})$, which is determined by the solute concentration distribution, as shown in Eq.~(\ref{B_with_ln(c)}). Consequently, the gap dynamics are coupled with solute transport, for which we derive a governing equation in Sec.~\ref{Subsec: Solute_transport}.

Finally, we note that in electrolyte solutions with unequal cation and anion diffusivities, the introduction of a concentration gradient spontaneously induces an electric field, $\tilde{E}_{\it induced}$, which arises to maintain the no-current condition~\cite{prieve1984motion,anderson1989colloid}.
For a symmetric binary ($z$:$z$) electrolyte with unequal cation and anion diffusion coefficients, the induced electric field is $\tilde{E}_{\it induced} = \beta_{D}\tilde{\varphi}_{T}\boldsymbol{\nabla}_{\Vert}\ln\tilde{c}$, where $\tilde{\varphi}_T=\tilde{k}_B\tilde{T}/z\tilde{e}$ is the thermal voltage and $\beta_D = (\tilde{D}_{+}-\tilde{D}_{-})/(\tilde{D}_{+}+\tilde{D}_{-})$ is the normalized diffusivity difference~\cite{prieve1984motion,anderson1989colloid,ault2024physicochemical}. 
Here, $\tilde{k}_B$ is the Boltzmann constant, $\tilde{T}$ is the absolute temperature, $\tilde{e}$ is the elementary charge, $z$ is the ion valence, and $\tilde{D}_{+}$ and $\tilde{D}_{-}$ are the diffusivities of the cations and anions, respectively.
The induced electric field gives rise to Maxwell stresses acting on the top plate~\cite{boyko2020nonuniform}.
Assuming that the permittivities of both the rigid plate and the surrounding air are negligible relative to that of the fluid, we neglect their contributions to the Maxwell stress and consider only the dielectric contribution of the fluid, given as $\tfrac{1}{2}\tilde{\varepsilon}\tilde{E}_{\it induced}^2$. Therefore, the resulting upward-directed dielectric force is given by
\begin{equation}
    \tilde{F}_{\it d}=\frac{\tilde{1}}{2}\tilde{\varepsilon}\int_0^{\tilde{\ell}}\tilde{E}_{\it induced}^2 d\tilde{x}= \frac{\tilde{1}}{2}\beta_D^2\tilde{\varepsilon}\tilde{\varphi}_T^2\tilde{\mathcal{M}} \quad \text{with} \quad     \tilde{\mathcal{M}}(\tilde{t}) = \int_0^{\tilde{\ell}}\left[\frac{\partial}{\partial\tilde{x}}\ln \left(\frac{\tilde{c}(\tilde{x},\tilde{z}=0,\tilde{t})}{\tilde{c}_L}\right)\right]^2d\tilde{x}.\label{F_d}
\end{equation}
While, in principle, the dielectric force $\tilde{F}_{\it d}$, Eq.~(\ref{F_d}), should be included in the force balance Eq.~(\ref{force_eq}), an evaluation using the physical parameters listed in Table~\ref{T1} for a representative microfluidic configuration reveals that the ratio between the dielectric and diffusioosmotic forces scales as  
$\beta_D^2\tilde{\varepsilon}\tilde{\varphi}_T^2 \tilde{h}_{i}^2 /(\tilde{\mu}\tilde{\Gamma}\tilde{\ell}^2) \approx 1.8\times 10^{-4}\beta_D^2$. Since $\beta_D$ is less than unity, we obtain that the dielectric contribution is negligible compared to the diffusioosmotic forcing. Moreover, for a symmetric binary electrolyte such as KCl, where the cation and anion have equal diffusivities~\cite{vanysek1993ionic}, $\beta_D = 0$, and consequently, the dielectric force $\tilde{F}_{\it d}$ vanishes. Therefore, in the subsequent analysis, we restrict our attention to the case of $\beta_D = 0$ and neglect the contribution of dielectric forces.

\subsection{Scaling analysis and non-dimensionalization of the evolution equation}\label{Subsec: Scaling}

Scaling by the characteristic dimensions, we introduce the non-dimensional variables
\begin{equation}
    h=\frac{\tilde{h}}{\tilde{h}_i}, \qquad x=\frac{\tilde{x}}{\tilde{\ell}}, \qquad z=\frac{\tilde{z}}{\tilde{h}_i}, \qquad 
    t = \frac{\tilde{t}}{\tilde{t}_c}, \qquad p = \frac{\tilde{p}}{\tilde{p}_c}, \qquad
    u_{\rm DOF}=\frac{\tilde{u}_{\rm DOF}}{\tilde{u}_c}, \qquad c=\frac{\tilde{c}}{\tilde{c}_L}, \qquad c_R=\frac{\tilde{c}_R}{\tilde{c}_L}, \label{ND_variebles_Our_Prob}
\end{equation}
where $\tilde{u}_c=\tilde{\Gamma}/\tilde{\ell}$ is the characteristic axial velocity based on the 
diffusioosmotic slip, $\tilde{p}_c=\tilde{\mu}\tilde{u}_{c}\tilde{\ell}/\tilde{h}_{i}^2$ is the characteristic lubrication pressure scale, and $\tilde{t}_c=(\tilde{\ell}/\tilde{h}_{i})^{3}(\tilde{\mu}/\tilde{k})$ is the characteristic time scale. We note that since the diffusioosmotic mobility $\tilde{\Gamma}$ scales inversely with the fluid viscosity $\tilde{\mu}$ \citep{prieve1984motion,anderson1989colloid}, the characteristic pressure based on the diffusioosmotic slip velocity is independent of the viscosity.

Substituting Eq.~(\ref{ND_variebles_Our_Prob}) into Eq.~(\ref{force_eq_sub}), we obtain the non-dimensional governing equation for the gap evolution
\begin{equation}
     \underset{\mathrm{Viscous\,resistance}}{\underbrace{ \frac{1}{h(t)^3}\frac{dh(t)}{dt}}} =\underset{\mathrm{Diffusioosmotic\,force}}{\underbrace{\frac{4}{27}\mathcal{F}_{\rm DOF}\frac{B(t)}{h(t)^2}}}+\underset{\mathrm{Elastic\,force}}{\underbrace{1-h(t)}} \quad \text{with} \quad B(t)=-\int_{0}^{1}\ln c dx+\frac{1}{2}\ln c_R,\label{ND_full_governing_eq}
\end{equation}
where we have introduced a non-dimensional parameter, referred to as the elasto-diffusioosmotic number,
\begin{equation}
\mathcal{F}_{\rm DOF}=\frac{81}{2}\frac{\tilde{t}_c}{\tilde{\ell}/\tilde{u}_c}=\frac{(81/2)\tilde{\mu}\tilde{\Gamma}/\tilde{h}_{i}^{2}}{\tilde{k}\tilde{h}_{i}/\tilde{\ell}}, \label{F_DOF}
\end{equation}
which represents the ratio between the diffusioosmotic driving forces and the elastic restoring forces.
In addition, the elasto-diffusioosmotic number $\mathcal{F}_{\rm DOF}$ can be interpreted as the ratio of the viscous--elastic to advection time scales.

The evolution equation (\ref{ND_full_governing_eq}) for the gap $h(t)$ represents the force balance between the viscous resistance, the diffusioosmotic force, and the restoring effect of the elastic force. 
For $\beta_D = 0$, corresponding to diffusioosmotic flow driven purely by the chemiosmotic contribution, both $\tilde{\Gamma}$ and the elasto-diffusioosmotic number $\mathcal{F}_{\rm DOF}$ are strictly positive. In contrast, the function $B(t)$ may take either positive or negative values, depending on the solute concentration profile.
Therefore, the direction of the diffusioosmotic forcing may vary over time. Since the function $B(t)$ depends on the solute concentration, we derive an additional governing equation for solute transport in the following subsection. Together, this set of equations provides a complete description of the time evolution of the gap.

\subsection{Advection--diffusion equation governing the solute transport}\label{Subsec: Solute_transport}

In this subsection, we derive a governing equation for solute transport, which should be solved together with Eq.~(\ref{ND_full_governing_eq}) to determine $h(t)$.
Using the non-dimensionalization Eq.~(\ref{ND_variebles_Our_Prob}), the advection--diffusion equation (\ref{adv-diff}) takes the form
\begin{equation}
   \epsilon^{2} \frac{\tilde{\ell}^{2}/\tilde{D}}{\tilde{t}_c}\frac{\partial c}{\partial t} + \epsilon^{2}\frac{\tilde{\ell}^2/\tilde{D}}{\tilde{\ell}/\tilde{u}_c}\left(u\frac{\partial c}{\partial x} + w\frac{\partial c}{\partial z}\right) = \epsilon^{2}\frac{\partial^{2} c}{\partial x^{2}}+\frac{\partial^{2} c}{\partial z^{2}}.\label{ND_conv_diff_divided}
\end{equation}
We recognize three relevant time scales in  Eq.~(\ref{ND_conv_diff_divided}).
These time scales involve the axial diffusive time scale $\tilde{\ell}^2/\tilde{D}$, the axial advection time scale $\tilde{\ell}/\tilde{u}_c$, and the viscous--elastic time scale $(\tilde{\ell}/\tilde{h}_{i})^{3}(\tilde{\mu}/\tilde{k})$, which we choose as the characteristic time scale for non-dimensionalization.

Using the definition of the elasto-diffusioosmotic number $\mathcal{F}_{\rm DOF}$ in Eq.~(\ref{F_DOF}), Eq.~(\ref{ND_conv_diff_divided}) can be expressed as 
\begin{equation}
  \frac{\epsilon^{2}}{\alpha}\frac{\partial c}{\partial t} + \frac{2\epsilon^{2}\mathcal{F}_{\rm DOF}}{81\alpha}\left(u\frac{\partial c}{\partial x} + w\frac{\partial c}{\partial z}\right) = \epsilon^{2}\frac{\partial^{2} c}{\partial x^{2}}+\frac{\partial^{2} c}{\partial z^{2}},\label{ND_adv_diff_with_Pe}
\end{equation}
where we have introduced a non-dimensional parameter, referred to as the elasto-diffusion number $\alpha$,
\begin{equation}
   \alpha=\frac{\tilde{t}_c}{\tilde{\ell}^2/\tilde{D}} = \frac{\tilde{\mu}\tilde{D}\tilde{\ell}}{\tilde{k}\tilde{h}_i^3},\label{alpha}
\end{equation}
which represents the the ratio of the viscous--elastic time scale $(\tilde{\ell}/\tilde{h}_{i})^{3}(\tilde{\mu}/\tilde{k})$ to the axial diffusive time scale $\tilde{\ell}^2/\tilde{D}$.
Recalling that $2\mathcal{F}_{\rm DOF}/81$ is the ratio of the viscous--elastic time scale $(\tilde{\ell}/\tilde{h}_{i})^{3}(\tilde{\mu}/\tilde{k})$ to the advection time scale $\tilde{\ell}/\tilde{u}_c$,
we note the ratio $2\mathcal{F}_{\rm DOF}/81\alpha$ appearing in the second term on the left-hand side of Eq.~(\ref{ND_adv_diff_with_Pe}) can be interpreted as the Péclet number $Pe=2\mathcal{F}_{\rm DOF}/81\alpha=\tilde{u}_c\tilde{\ell}/\tilde{D}=\tilde{\Gamma}/\tilde{D}$, representing the ratio of the axial diffusive time scale $\tilde{\ell}^2/\tilde{D}$ to the advection time scale $\tilde{\ell}/\tilde{u}_c$. We note that the Péclet number is typically an order-one parameter, i.e., $Pe = O(1)$~\cite{ault2018diffusiophoresis,teng2023diffusioosmotic}.

The governing equation (\ref{ND_adv_diff_with_Pe}) is supplemented by no-flux boundary conditions at the bottom and top surfaces, given in Eq.~(\ref{D_bc_conc}), which, using Eq.~(\ref{ND_variebles_Our_Prob}), can be expressed in a non-dimensional form as 
\begin{equation}
    \left.\frac{\partial c}{\partial z}\right|_{z=0}=0 \quad \text{and} \quad \left.\frac{\partial c}{\partial z}\right|_{z=h(t)}=\frac{\epsilon^{2}}{\alpha}\frac{dh(t)}{dt}\left.c\right|_{z=h(t)}.\label{ND_bc_conc}
\end{equation}
Assuming that $\alpha= O(1)$ and $Pe=O(1)$, we are interested in the leading-order dynamics through a formal expansion with the small parameter $\epsilon=\tilde{h}_{i}/\tilde{\ell}\ll1$. As can be seen in the non-dimensional governing equation (\ref{ND_adv_diff_with_Pe}) and the boundary conditions (\ref{ND_bc_conc}), only $\epsilon^{2}$ appears. Thus, we seek the solution in the form
\begin{equation}
c=c_0+\epsilon^{2}c_{1}+O(\epsilon^{4}).\label{concentration_solution_form}
\end{equation}
At the leading order, $O(1)$, we obtain
\begin{equation}
    \frac{\partial^{2}c_0}{\partial z^2}=0 \quad  \text{subject to} \quad \left.\frac{\partial c_{0}}{\partial z}\right|_{z=0}=0 \quad  \text{and} \quad \left.\frac{\partial c_{0}}{\partial z}\right|_{z=h(t)}=0.\label{leading_order}
\end{equation}
Integrating the governing equation and applying the no-flux boundary conditions, we find that $c_0=c_0(x,t)$, i.e., the solute concentration is independent of $z$ to the leading order in $\epsilon$.

At the next order, $O(\epsilon^{2})$, the governing equation (\ref{ND_adv_diff_with_Pe}) takes the form
\begin{equation}
    \frac{1}{\alpha}\frac{\partial c_0}{\partial t} + \frac{2\mathcal{F}_{\rm DOF}}{81\alpha}u\frac{\partial c_0}{\partial x} = \frac{\partial^{2}c_0}{\partial x^2}+\frac{\partial^{2}c_1}{\partial z^2},\label{main_eq_O(eps^2)}
\end{equation}
and the boundary conditions become
\begin{equation}
    \left.\frac{\partial c_1}{\partial z}\right|_{z=0} \quad \text{and} \quad\left.\frac{\partial c_1}{\partial z}\right|_{z=h(t)}=\frac{1}{\alpha}\frac{dh(t)}{dt}c_0|_{z=h(t)}.\label{bc_O(eps2)}
\end{equation}
Recalling that $c_0=c_0(x,t)$, we integrate Eq.~(\ref{main_eq_O(eps^2)}) with respect to $z$ from 0 to $h(t)$ and use the boundary conditions (\ref{bc_O(eps2)}) to eliminate the term involving $c_1$ in Eq.~(\ref{main_eq_O(eps^2)}) and obtain
\begin{equation}
    \frac{\partial c_{0}}{\partial t}h(t) + \frac{2}{81}\mathcal{F}_{\rm DOF}\frac{\partial c_0}{\partial x}q(x,t)=\alpha\frac{\partial^{2}c_0}{\partial x^2}h(t) +\frac{dh(t)}{dt}c_0.\label{governing_adv_diff_leading_order}
\end{equation}
Here, $q(x,t)=\int_{0}^{h(t)}udz$ is the dimensionless flow rate obtained from Eq.~(\ref{D_flow_rate}) as
\begin{equation}
    q(x,t)= \frac{\tilde{\ell}/\tilde{u}_c}{\tilde{t}_c}\frac{d h(t)}{dt}\left(\frac{1}{2}-x \right)-\frac{1}{2}\frac{\tilde{\Gamma}/\tilde{\ell}}{\tilde{u}_c}h(t)\ln c_R=\frac{81}{2\mathcal{F}_{\rm DOF}}\frac{d h(t)}{dt}\left(\frac{1}{2}-x \right)-\frac{1}{2}h(t)\ln c_R,\label{q_with_const}
\end{equation}
where in the last equality in Eq.~(\ref{q_with_const}) we have used the fact that $2\mathcal{F}_{\rm DOF}/81=\tilde{t}_c /(\tilde{\ell}/\tilde{u}_c)$ and  $\tilde{u}_c=\tilde{\Gamma}/\tilde{\ell}$.

Finally, substituting Eq.~(\ref{q_with_const}) into Eq.~(\ref{governing_adv_diff_leading_order}), the advection--diffusion equation reduces to
\begin{equation}
  \frac{\partial c}{\partial t}h(t) + \frac{\partial c}{\partial x}\left[\frac{dh(t)}{dt}\left(\frac{1}{2}- x\right)-\frac{1}{81}\mathcal{F}_{\rm DOF}h(t)\ln c_R\right]=   \alpha\frac{\partial^{2}c}{\partial x^2}h(t) +\frac{dh(t)}{dt}c,\label{adv_diff_with final}
\end{equation}
where we dropped the subscript ``0" in $c(x,t)$ for simplicity.

The evolution equations (\ref{ND_full_governing_eq}) and (\ref{adv_diff_with final}) form a set of two-way coupled nonlinear equations, governing the fluid--structure interaction driven by diffusioosmotic flow, that should be solved at once to determine both the gap height $h(t)$ and solute concentration $c(x,t)$.~The governing equations (\ref{ND_full_governing_eq}) and (\ref{adv_diff_with final}) are supplemented by the boundary conditions, corresponding to fixed solute concentrations at the left and right edges,
\begin{equation}
     c(x=0,t)=1 \quad \text{and}  \quad c(x=1,t)=c_R,\label{BCs on c ND}
\end{equation}
and the initial conditions,
\begin{equation}
     c(x,t=0)=1 \quad \text{and}  \quad h(t=0)=1.\label{ICs on c and h ND}
\end{equation}

We observe that the governing equations (\ref{ND_full_governing_eq}) and (\ref{adv_diff_with final}) and the boundary conditions (\ref{BCs on c ND}) depend on three key dimensionless parameters: $\mathcal{F}_{\rm DOF}$, $\alpha$, and $c_R$. 
The parameters $\mathcal{F}_{\rm DOF}$ and $\alpha$ remain constant for a given configuration, whereas $c_R$ may vary by changing the solute concentration at the right edge. Therefore, in the next section, we mainly present the results for fixed values of $\mathcal{F}_{\rm DOF} = 40.5$ and $\alpha = 1$, while varying $c_R$. These values of $\mathcal{F}_{\rm DOF}$ and $\alpha$ correspond to the Péclet number $Pe=2\mathcal{F}_{\rm DOF}/81\alpha=\tilde{\Gamma}/\tilde{D}=1$.

\section{Results}\label{Sec: Results}

In this section, we present our theoretical results for the fluid--structure interaction dynamics and the onset of fluid--structure instability driven by diffusioosmotic flow. We further validate the predictions
of our reduced-order theoretical model against the two-dimensional numerical simulations with the finite-element software COMSOL Multiphysics (version 6.2, COMSOL AB, Stockholm, Sweden).

\subsection{Steady-state results}

First, in this subsection, we examine the steady-state response of the system. The governing equations for the gap height evolution, Eq. (\ref{ND_full_governing_eq}), and the solute transport, Eq.~(\ref{adv_diff_with final}), at steady state take the form
\begin{subequations}
\begin{gather}
\frac{4}{27}\mathcal{F}_{\rm DOF}\frac{B_{\rm ss}}{h_{\rm ss}^2} + 1 - h_{\rm ss} = 0, \qquad
B_{\rm ss} = -\int_{0}^{1}\ln c_{\rm ss}(x)\, dx + \frac{1}{2}\ln c_R, \label{ND_governing_eq_h_ss_1} \\
\alpha\frac{d^{2}c_{\rm ss}}{d x^2}  +\frac{1}{81}\mathcal{F}_{\rm DOF}\ln c_R\frac{d c_{\rm ss}}{d x} = 0 
\quad \Rightarrow \quad
\frac{d^{2}c_{\rm ss}}{d x^2}  +\frac{1}{2}Pe\ln c_R\frac{d c_{\rm ss}}{d x} = 0. \label{adv_diff_gv_eq_ss_1}
\end{gather}\label{sub_ep_ss}\end{subequations}
where the subscript $\rm ss$ denotes the steady-state solution. 

At steady state, the governing equations (\ref{sub_ep_ss}) become one-way coupled, allowing for an analytical solution of the advection--diffusion equation (\ref{adv_diff_gv_eq_ss_1}), given as  
\begin{equation}
c_{\rm ss}(x)= \frac{1}{ c_R^{Pe/2}-1} \left[\left(1 - c_R\right) c_R^{(Pe/2)(1-x)} + c_R^{1 + (Pe/2)}-1\right].\label{css_Analytic}
\end{equation}
Knowing the steady-state concentration $c_{\rm ss}(x)$, we can find the parameter $B_{\rm ss}$ and the steady-state gap height $h_{\rm ss}$ using Eq.~(\ref{ND_governing_eq_h_ss_1}). We note that the steady-state concentration distribution and the parameter $B_{\rm ss}$ are independent of the steady-state gap height $h_{\rm ss}$.

Once $c_{\rm ss}(x)$ and $h_{\rm ss}$ are determined, the corresponding steady-state pressure distribution in a non-dimensional form can be calculated using Eq.~(\ref{D_pressure}) as 
\begin{equation}
    p_{\rm ss}(x) = -\frac{6}{h_{\rm ss}^2}\left(\ln c_{\rm ss}(x) - x\ln c_R\right).\label{pss_Analytic}
\end{equation}
\begin{figure}[t]
    \centering
    \includegraphics[scale=1.2]{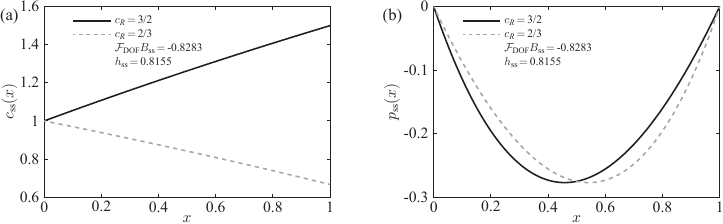}
    \caption{Steady-state concentration and pressure distributions in the fluid--structure interaction driven by diffusioosmotic flow. (a) The concentration distribution at steady state, given by Eq.~(\ref{css_Analytic}). (b) The corresponding pressure distribution at steady state, given by Eq.~(\ref{pss_Analytic}). Black solid and gray dashed curves represent the results for $c_R=3/2$ and $c_R=2/3$, respectively. The pressure profiles exhibit mirror symmetry across the channel center, consistent with the inverse ratios of $c_R$. The steady-state gap height for both cases is $h_{\rm ss}=0.8155$. All calculations were performed using $\alpha=1$ and $\mathcal{F}_{\rm DOF}=40.5$.}
    \label{F2}
\end{figure}

Figure~\ref{F2}(a,b) presents the steady-state concentration and pressure distributions for $c_R=3/2$ and $c_R=2/3$. Due to the non-negligible advective contribution with a Péclet number of $Pe=1$, the concentration distributions shown in Fig.~\ref{F2}(a) slightly deviate from a linear behavior. We observe in Fig.~\ref{F2}(b) that the diffusioosmotic flow induces a negative fluidic pressure, which acts as a destabilizing force. This negative diffusioosmotic pressure reduces the gap height until it reaches a stable steady-state value, $h_{\rm ss}$, where the elastic restoring force balances the diffusioosmotic driving force. Furthermore, consistent with the inverse ratio of the $c_R$ values, the pressure distributions exhibit mirror symmetry about the channel center, and the steady-state gap height $h_{\rm ss}=0.8155$ is identical in both cases.

Scaling analysis based on lubrication theory shows that the negative fluidic pressure generated by diffusioosmotic flow, represented by the first term on the right-hand side of Eq.~(\ref{ND_full_governing_eq}), scales inversely with the gap height, $h$. Therefore, as the gap height decreases due to negative fluidic pressure, the pressure becomes increasingly negative, further amplifying the reduction in gap height. In the next subsection, we show that above a certain critical threshold of the concentration at the right edge, $c_R=c_{R, \it CR}$, this mechanism gives rise to the onset of the fluid--structure instability, ultimately leading to the collapse of the top plate onto the bottom surface.

\subsection{Dynamic results}

In this subsection, we investigate the fluid--structure interaction dynamics and the onset of fluid--structure instability by solving numerically the gap evolution equation (\ref{ND_full_governing_eq}) coupled with the advection--diffusion equation (\ref{adv_diff_with final}) using MATLAB's routine \texttt{ode15s}. We provide the details of the numerical procedure in Appendix~\ref{AppA}.

First, we present in Fig.~\ref{F3}(a,b) the time evolution of $\mathcal{F}_{\rm DOF}B(t)$ and the corresponding gap height $h(t)$ for different values of $c_R$. The sign of $\mathcal{F}_{\rm DOF}B(t)$ determines the direction of the diffusioosmotic forcing (see Eq.~(\ref{ND_full_governing_eq})). Positive values 
of $\mathcal{F}_{\rm DOF}B(t)$ correspond to a rise of the rigid top plate, whereas negative values of $\mathcal{F}_{\rm DOF}B(t)$ indicate its descent.
Figure~\ref{F3}(a) clearly shows that at early times, $\mathcal{F}_{\rm DOF}B(t)$ is positive, but at late times, it becomes negative for all values of $c_R$. Thus, except for the small increase in the gap height observed at early times, we identify in Fig.~\ref{F3}(b) three distinct dynamic behaviors at late times, all characterized by a decrease in the gap height $h(t)$.
(i)~Below the threshold value of $c_{R,{\it CR}}=1.5617$, corresponding to $\mathcal{F}_{\rm DOF}B(t)=-1$~(see Sec.~\ref{Results_Section3} and Fig.~\ref{F6}), the top plate approaches a stable steady-state height. (ii) However, a small increase in $c_R$ above the threshold value (here, we show an increase by $0.1\%$) triggers the onset of the fluid--structure instability, leading to a drastic change in the behavior of the system. In this regime, the gap height evolution exhibits a bottleneck, during which the descent of the top plate slows down. However, the top plate eventually accelerates and collapses onto the bottom surface. (iii) As $c_R$ is further increased, the bottleneck behavior disappears, and the top plate collapses immediately onto the bottom surface.
Therefore, we refer to these three distinct dynamic behaviors as the \emph{stable}, \emph{bottleneck}, and \emph{immediate collapse} dynamic regimes.

It is evident from Fig.~\ref{F3}(b) that in both the bottleneck and immediate collapse regimes, the top plate significantly slows down just before approaching the bottom surface. We rationalize this slowdown by noting that at late times, when 
$h\ll1$, the diffusioosmotic force balances the viscous resistance in Eq.~(\ref{ND_full_governing_eq}), while the elastic force becomes negligible. Since the parameter $\mathcal{F}_{\rm DOF}B(t)$ remains approximately constant at late times, exhibiting only weak time dependence as shown in Fig.~\ref{F3}(a), the diffusioosmotic force scales as $h^{-2}$, while the viscous resistance scales as $h^{-3}(dh/dt)$. This balance leads to an exponential approach of the top plate toward the bottom surface.

We note that~\citet{boyko2020nonuniform}, who studied the fluid--structure interaction driven by non-uniform electroosmotic flow, reported three qualitatively similar dynamic behaviors. This similarity in behavior is due to the similar scaling of the electroosmotic and diffusioosmotic forcings, both of which scale as $h^{-2}$. However, there is an important difference. While for non-uniform electroosmotic flow the dynamics of the gap height is governed by a single evolution equation for $h(t)$, in the case of diffusioosmotic flow, the dynamics of the gap height is governed by coupled evolution equations for both $h(t)$ and the solute concentration $c(x,t)$.

\begin{figure}[t]
    \centering
    \includegraphics[scale=1.2]{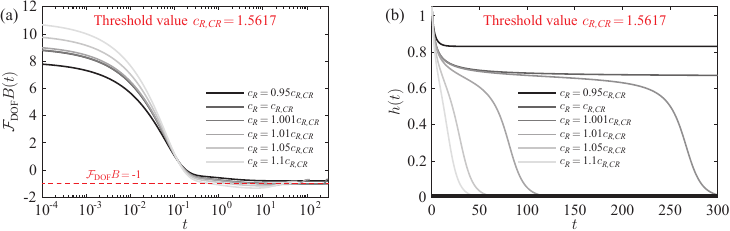}
    \caption{Dynamic results of the fluid--structure interaction driven by diffusioosmotic flow. (a) The time evolution of $\mathcal{F}_{\rm DOF}B(t)$ for various values of $c_R$. (b) The corresponding time evolution of the gap height $h(t)$ for various values of $c_R$, indicating three different dynamic behaviors. Below the threshold value of $c_{R,{\it CR}} = 1.5617$, corresponding to $\mathcal{F}_{\rm DOF}B(t)=-1$, the top plate approaches a stable steady-state height. Increasing $c_R$ slightly above the threshold value (we show an increase by $0.1\%$) leads to the onset of the fluid--structure instability characterized by a bottleneck. For larger values of $c_R$ (e.g., $c_R=1.1c_{R,{\it CR}}=1.7179$), the bottleneck disappears and the top plate immediately collapses onto the bottom surface. Note that in panel (a) the curves of $\mathcal{F}_{\rm DOF}B(t)$ for $c_R=c_{R,{\it CR}}$ and $c_R=1.001c_{R,{\it CR}}$ are almost indistinguishable. All calculations were performed using $\alpha=1$ and $\mathcal{F}_{\rm DOF}=40.5$.}\label{F3}
\end{figure}

\begin{figure}[t]
    \centering
\includegraphics[scale=1.2]{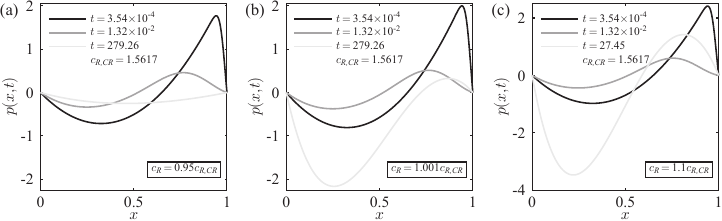}
    \caption{Time evolution of the pressure distribution in the fluid--structure interaction driven by diffusioosmotic flow.  (a,b,c) The time evolution of the pressure distribution corresponding to the (a) stable, (b) bottleneck, and (c) immediate collapse dynamic regimes. Black, dark gray, and light gray curves represent times $t=3.54\times10^{-4}$, $ 1.32\times 10^{-2}$, and $279.26$, respectively, except for panel (c), where the light gray curve corresponds to $t=27.45$. All calculations were performed using $\alpha=1$ and $\mathcal{F}_{\rm DOF}=40.5$.}
    \label{F4}
\end{figure}

To provide further insight into the dynamic behavior of the system, we present in Fig.~\ref{F4} the time evolution of the non-dimensional pressure distribution for three different values of $c_R$. Specifically, Figs.~\ref{F4}(a), \ref{F4}(b), and \ref{F4}(c) correspond to the stable ($c_R=0.95c_{R,{\it CR}}$), bottleneck ($c_R=1.001c_{R,{\it CR}})$, and immediate collapse ($c_R=1.1c_{R,{\it CR}}$) dynamic regimes, respectively. Using Eqs.~(\ref{D_pressure}), (\ref{D_A}), (\ref{ND_variebles_Our_Prob}), and (\ref{F_DOF}), the dimensionless pressure distribution $p(x,t)$ reads 
\begin{equation}
    p(x,t) = \frac{243}{\mathcal{F}_{\rm DOF}}\frac{1}{h(t)^3}\frac{dh(t)}{dt}x(x-1)-\frac{6}{h(t)^2}\left(\ln c(x,t) - x\ln c_R\right),\label{ND_pressure}
\end{equation}
where $h(t)$ and $c(x,t)$ are obtained from the solution of Eqs.~(\ref{ND_full_governing_eq}) and (\ref{adv_diff_with final}).

At early times, the pressure distributions are qualitatively similar for all dynamic regimes. However, at later times, we observe a distinct shift in behavior. While the pressure distribution in the stable regime generates a moderate downward fluidic force that is balanced by the elastic restoring force at late times (Fig.~\ref{F4}(a)), the pressure distributions in the bottleneck and immediate collapse regimes create significantly stronger downward fluidic forces (Fig.~\ref{F4}(b,c)), ultimately driving the collapse of the top plate. In particular, within the immediate collapse regime, the late-time pressure distribution generates an even stronger downward fluidic force than that observed in the bottleneck regime, resulting in a more rapid collapse of the top plate.

\begin{figure}[b]
    \centering
\includegraphics[scale=1.2]{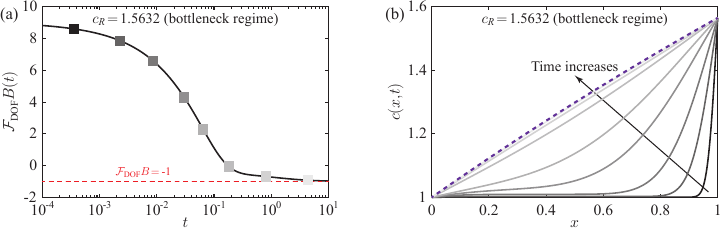}
    \caption{Dynamic results of the fluid--structure interaction driven by diffusioosmotic flow for the bottleneck regime with $c_R=1.001c_{R,{\it CR}}=1.5632$. (a) The time evolution of $\mathcal{F}_{\rm DOF}B(t)$, where the gray-scale squares mark specific times. (b) The corresponding time evolution of the concentration distribution, illustrated by gray-scale curves. The purple dashed line represents the analytical steady-state solution, given by Eq.~(\ref{css_Analytic}). All calculations were performed using $\alpha=1$ and $\mathcal{F}_{\rm DOF}=40.5$.}
    \label{F5}
\end{figure}

In addition to the time evolution of the gap height, it is of particular interest to study the time evolution of the solute concentration, which drives the diffusioosmotic flow and the resulting fluid--structure interaction.
Figure \ref{F5}(a) presents the time evolution of $\mathcal{F}_{\rm DOF}B(t)$ for the bottleneck regime with $c_R=1.001c_{R,{\it CR}}$, where gray-scale squares mark specific times. In Fig.~\ref{F5}(b), we present the corresponding time evolution of the concentration distribution.
Although in the bottleneck regime the gap height $h(t)$ continues to decrease until the top plate collapses onto the bottom surface, without satisfying the steady-state equation (\ref{ND_governing_eq_h_ss_1}), it is evident from Fig.~\ref{F5}(b) that the concentration distribution approaches its steady-state profile, given by Eq.~(\ref{css_Analytic}). More importantly, the concentration distribution reaches a steady state when $t \approx O(1)$, which is significantly earlier than the time at which the top plate approaches the bottom surface (see Fig.~\ref{F3}(b)).
This result is consistent with Fig.~\ref{F3}(a), which shows that the parameter $\mathcal{F}_{\rm DOF}B(t)$, directly related to the concentration (see Eq.~(\ref{ND_full_governing_eq})), exhibits only weak time dependence once $t\gtrsim O(1)$.

\subsection{Linear stability analysis and investigation of the instability onset}\label{Results_Section3}

In the previous subsection, we demonstrated the existence of fluid--structure instability driven by diffusioosmotic flow, which occurs when the concentration $c_R$ exceeds a critical threshold $c_{R, \it CR}$. In this subsection, we perform a linear stability analysis of the system to gain deeper insight into the physical mechanisms underlying the fluid--structure instability. 

We recall that, as shown in Fig.~\ref{F5}(b), the concentration rapidly evolves towards the steady-state distribution, given by Eq.~(\ref{css_Analytic}), which is independent of the gap height. As a result, the parameter $\mathcal{F}_{\rm DOF}B(t)$ remains approximately constant once $t\gtrsim O(1)$, exhibiting weak time dependence, as shown in Fig.~\ref{F3}(a). In particular, the curves of $\mathcal{F}_{\rm DOF}B(t)$ for $c_R=c_{R,{\it CR}}$ and $c_R=1.001c_{R,{\it CR}}$ are almost indistinguishable and remain constant for $t\gtrsim O(1)$.

Therefore, we consider a small perturbation~\emph{only} of the gap height from its equilibrium value $h_{\rm ss}$, while assuming that the value of $B$ remains constant and equal to its steady-state value $B_{\rm ss}$, thereby eliminating the need to introduce a perturbation to the concentration. 
To this end, we let $h(t) = h_{\rm ss}\left(1+\epsilon_0 e^{\sigma t}\right)$, where $\epsilon_0\ll1$ is some small perturbation and $\sigma$ is the non-dimensional growth rate.
Substituting this expression into Eq.~(\ref{ND_full_governing_eq}), the leading order provides the equation for the steady-state solution $h_{\rm ss}$
\begin{equation}
    \frac{4}{27}\mathcal{F}_{\rm DOF}\frac{B_{\rm ss}}{h_{\rm ss}^2} + 1 -h_{\rm ss}=0.\label{Leading_order_h}
\end{equation}
At the first order, $O(\epsilon_0)$, we obtain the equation for the growth rate  $\sigma$
\begin{equation}
    \sigma = -\frac{8}{27}\mathcal{F}_{\rm DOF}B_{\rm ss}-h_{\rm ss}^3 = h_{\rm ss}^2\left(2-3h_{\rm ss}\right),\label{First_order_h}
\end{equation}
indicating that the system may become unstable for certain negative values of $\mathcal{F}_{\rm DOF}B_{\rm ss}$. From Eq.~(\ref{First_order_h}), it follows that the critical steady-state height $h_{{\rm ss},{\it CR}}$, below which the system becomes unstable, and the corresponding threshold value $\mathcal{F}_{\rm DOF}B_{{\rm ss,{\it CR}}}$, are given by 
\begin{equation}
    h_{{\rm ss},{\it CR}} = \frac{2}{3} \quad \text{and} \quad \mathcal{F}_{\rm DOF}B_{{\rm ss,{\it CR}}}=-1.\label{critical_values}
\end{equation}

\begin{figure}
    \centering
\includegraphics[scale=1.2]{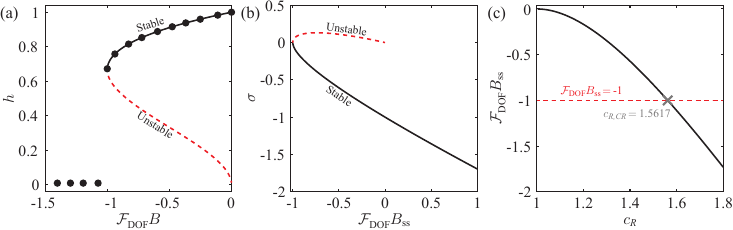}
    \caption{Steady-state analysis of the fluid--structure interaction driven by diffusioosmotic flow. (a) The steady-state gap height $h$ as a function of $\mathcal{F}_{\rm DOF}B$. The black solid curve represents the stable steady-state solution, while the red dashed curve represents the unstable result obtained from the linear stability analysis. Black dots represent the results of the dynamic simulation for different values of $c_R$, where we report the minimum value of $B$ in the simulation. At $\mathcal{F}_{\rm DOF}B =-1$, the stable steady-state solution intersects an unstable solution and disappears at a saddle-node bifurcation. (b) The growth rate $\sigma$ as a function of $\mathcal{F}_{\rm DOF}B_{\rm ss}$ obtained from the linear stability analysis. The black solid curve corresponds to the stable steady-state solution and the red dashed curve corresponds to the unstable result in panel (a).
    (c) The parameter $\mathcal{F}_{\rm DOF}B_{\rm ss}$ as a function of $c_R$. The intersection of the curve with $\mathcal{F}_{\rm DOF}B_{\rm ss}=-1$ (red dashed line) occurs at $c_{R,{\it CR}}=1.5617$ (gray cross), indicating the threshold value of $c_R$ for the onset of the fluid--structure instability.
    All calculations were performed using $\alpha=1$, $\mathcal{F}_{\rm DOF}=40.5$, and $c_R>1$.}
    \label{F6}
\end{figure}

Figure~\ref{F6}(a) presents the steady-state gap height as a function of $\mathcal{F}_{\rm DOF}B$. The black solid curve represents the linearly stable steady-state solution of Eq.~(\ref{Leading_order_h}), whereas the red dashed curve represents the linearly unstable steady-state solution. These two solutions coincide and disappear at a saddle-node bifurcation (see, e.g.,~\cite{gomez2018delayed,boyko2020nonuniform}), occurring at $\mathcal{F}_{\rm DOF}B_{{\rm ss,{\it CR}}}=-1$ with $h_{\rm ss,\it CR}=2/3$, consistent with Eq.~(\ref{critical_values}). Black dots represent the predictions of the dynamic theoretical model, showing excellent agreement with the results
of linear stability analysis. In the results presented here, we use the minimum value of $B$ obtained in the simulation.
Figure~\ref{F6}(b) illustrates the growth rate $\sigma$ as a function of $\mathcal{F}_{\rm DOF}B_{\rm ss}$, given by Eq.~(\ref{First_order_h}). As expected, the growth rate, represented by the black solid curve and corresponding to the linearly stable steady-state solution for $h_{\rm ss}$, is negative. Similarly, the growth rate, represented by the red dashed curve and corresponding to the linearly unstable steady-state solution for the gap height, is positive. For $\mathcal{F}_{\rm DOF}B_{\rm ss}=-1$, the growth rate is zero, indicating the threshold of fluid--structure instability.

Aiming to rationalize the dynamic predictions of the theoretical model and the results of the linear stability analysis for the onset of instability, we present in Fig.~\ref{F6}(c) the parameter $\mathcal{F}_{\rm DOF}B_{\rm ss}$ as a function of $c_R$. We observe that the intersection between the black curve and the red dashed line $\mathcal{F}_{\rm DOF}B_{\rm ss}=-1$ occurs at $c_{R, \it CR}=1.5617$ (gray cross), indicating the onset of instability, consistent with the result of the linear stability analysis, given by Eq.~(\ref{critical_values}). This finding provides a theoretical justification for the threshold value $c_{R, \it CR}=1.5617$, previously identified through dynamic simulations shown in Fig.~\ref{F3}. 

\begin{figure}[t]
    \centering
\includegraphics[scale=1.2]{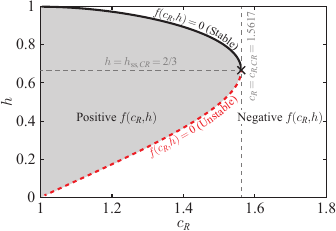}
    \caption{Stability diagram showing the contour of $f(c_R,h)$ for different values of $h$ and $c_R$, given by Eq. (\ref{f_stability_state}). The positive (gray) region of $f$ represents the dominance of the restoring elastic effect, whereas the negative (white) region represents the dominance of the diffusioosmotic force. The black solid curve represents the zero contour of the function $f(c_R,h)$, which corresponds to the linearly stable steady-state solution for $h_{\rm ss}$. The red dashed curve represents the zero contour of the function $f(c_R,h)$, which corresponds to the linearly unstable steady-state solution. The black cross marks the point $c_{R, {\it CR}}=1.5617$ and $h_{\rm ss,{\it CR}}=2/3$, indicating the threshold values of concentration and gap height for the onset of fluid--structure instability.
    All calculations were preformed using $\alpha=1$ and $\mathcal{F}_{\rm DOF}=40.5$.}
    \label{F7}
\end{figure}

To better understand the behavior of the system and its transition to fluid--structure instability, we define the function $f(c_R,h)$, representing the combination of the diffusioosmotic and elastic forces,
\begin{equation}
        f(c_R,h) =\underset{\mathrm{Negative\,diffusioosmotic\,force}}{\underbrace{\frac{4}{27}\mathcal{F}_{\rm DOF}\frac{B_{\rm ss}(c_R)}{h^2}}}+\underset{\mathrm{Positive\,elastic\,force}}{\underbrace{1-h}},\label{f_stability_state}
\end{equation}
where its zero contour, $f(c_R,h)=0$, corresponds to the equation for the steady-state solution $h_{\rm ss}$, as given by Eq.~(\ref{Leading_order_h}).
We note that from Eq.~(\ref{ND_governing_eq_h_ss_1}), it follows that the parameter $B_{\rm ss}$ solely depends on the concentration at the right edge, $c_R$, and the Péclet number, $Pe$, but is independent of the gap height.

We present in Fig.~\ref{F7} a stability diagram showing the contour of $f(c_R,h)$ for different values of $h$ and $c_R$.
Black solid and red dashed curves represent the zero contour of the function $f(c_R,h)$, where the negative diffusioosmotic force balances the positive restoring elastic force. The black solid curve represents the linearly stable steady-state solution for $h_{\rm ss}$, whereas the red dashed curve represents the linearly unstable steady-state solution, consistent with the results shown in Fig.~\ref{F6}(a). The black cross marks the point $c_{R, \it CR}=1.5617$ and $h_{\rm ss,\it CR}=2/3$, indicating the threshold values for the onset of fluid--structure instability. As the diffusioosmotic force at steady state is always negative (see Fig.~\ref{F6}(c)) and the elastic force remains positive for 
$h<1$, the positive (gray) region of $f$ indicates the dominance of the restoring elastic effect, whereas the negative (white) region reflects the dominance of the diffusioosmotic force.

Furthermore, we identify several distinct behaviors depending on the values of $c_R$, $h$, and the sign of $f$. 
When $c_R>c_{R, \it CR}$, the function $f$ is always negative, indicating that the diffusioosmotic downward force consistently dominates the elastic restoring force, leading to the collapse of the top plate onto the bottom surface for any gap height $h$. This behavior corresponds to the bottleneck and immediate collapse regimes, shown in Fig.~\ref{F3}(b).

However, for $c_R<c_{R, \it CR}$, the behavior of the system depends strongly on the value of $h$ and the sign of $f$: (i) When $h>h_{\rm ss,\it CR}$ and $f<0$, the dominant diffusioosmotic downward force causes the top plate to descend until it reaches a stable steady-state height, represented by the black solid curve. This behavior corresponds to the stable regime, shown in Fig.~\ref{F3}(b).
(ii) Similarly, for $f>0$, the dominant restoring elastic force causes the plate to rise toward the stable steady-state height, even when the gap height is below the critical value $h_{\rm ss,\it CR}$. (iii) In contrast, when $h<h_{\rm ss,\it CR}$ and $f<0$, fluid--structure instability exists, and the dominant diffusioosmotic downward force causes the collapse of the top plate onto the bottom surface. 

\subsection{Finite-element numerical validation}\label{Finite-element sim}

To validate the predictions of our theoretical model, we carry out finite-element numerical simulations with the commercial software COMSOL Multiphysics (version 6.2, COMSOL AB, Stockholm, Sweden). In Appendix~\ref{AppB}, we provide detailed descriptions of the governing equations, boundary conditions, domain discretization, and physical parameters used in the finite-element simulations.

\begin{figure}[t]
    \centering
    \includegraphics[scale=1.2]{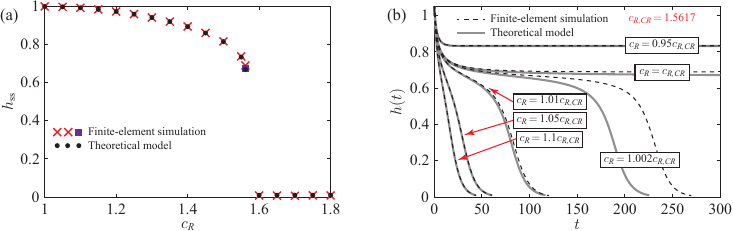}
    \caption{Comparison of finite-element simulation results and theoretical model predictions of the fluid--structure interaction driven by diffusioosmotic flow. 
    (a) The steady-state gap height, $h_{\rm ss}$, as a function of $c_R$. Black dots represent the theoretical model predictions. Red crosses and a purple square represent the finite-element simulation results. 
    (b) The time evolution of the gap height, $h(t)$, for different values of $c_R$. Gray solid lines represent the theoretical model predictions and black dashed lines represent the finite-element simulation results. All calculations were performed using the values from Table~\ref{T1}, with $\alpha=1$ and $\mathcal{F}_{\rm DOF}=40.5$.}
    \label{F8}
\end{figure}

Figure \ref{F8} presents a comparison of finite-element simulation results and theoretical model predictions. Figure~\ref{F8}(a) shows the steady-state gap height, $h_{\rm ss}$, as a function of $c_R$, showing excellent agreement between the theoretical predictions (black dots) and the finite-element simulation results (red crosses). Specifically, the threshold value of $c_R$ for the onset of fluid--structure instability, $c_{R,\it CR}$, is also well predicted. In the finite-element simulations, the threshold value is $c_{R,\it CR,\it FE}=1.5628$ (purple square), which is slightly higher than the theoretical prediction of $c_{R, CR}=1.5617$.

Next, in Fig.~\ref{F8}(b) we present the time evolution of the gap height $h(t)$ for different values of $c_R$. Gray solid lines represent the theoretical model predictions and black dashed lines represent the finite-element simulation results. 
We observe that the theoretical model accurately captures the transient dynamics of the gap height, showing excellent agreement with the finite-element simulation results, except in the close vicinity of the threshold value of $c_R$.
For $c_R=c_{R, \it CR}=1.5617$ and for a slightly higher value such as $c_R=1.002c_{R, \it CR}$, a small discrepancy arises between the theoretical model predictions and the finite-element simulation results at late times. This discrepancy is due to the small difference in the threshold value $c_{R, CR}$.
As a result, for $c_R=1.002c_{R, \it CR}$, the theoretical model underestimates the total collapse time of the top plate.

\section{Concluding remarks}\label{CR}

In this work, we studied the fluid--structure interaction and instability of a thin electrolyte film confined between a rigid surface and an elastic substrate driven by diffusioosmotic flow. Applying the lubrication approximation and modeling the elastic substrate as a rigid plate connected to a linear spring, we derived a pair of two-way coupled nonlinear governing equations for the evolution of the film thickness $h(t)$ and the solute concentration $c(x,t)$.
Our theoretical analysis revealed that, upon application of a certain solute concentration difference between the left and right edges, negative pressures induced by diffusioosmotic flow lead to the onset of fluid--structure instability.
Through dynamic simulations of the coupled governing equations, we identified the critical physical parameters for the onset of fluid--structure instability, along with three distinct dynamic regimes: (i) a stable steady state, (ii) a bottleneck, and (iii) an immediate collapse. Furthermore, we performed a linear stability analysis of the system to rationalize the critical values of the film thickness
$h$ and the right-edge concentration $c_R$ for the onset of instability predicted by the dynamic simulations.
Finally, we validated our theoretical predictions with finite-element numerical simulations and found excellent agreement.

It is worth noting that throughout this work, we presented results for a positive diffusioosmotic mobility $\tilde{\Gamma}$, corresponding to diffusioosmotic flow arising solely from the chemiosmotic contribution. In this case, we demonstrated the existence of fluid--structure instability above a certain threshold of $c_R$.
However, when both chemiosmotic and electroosmotic contributions drive diffusioosmotic flow, the diffusioosmotic mobility $\tilde{\Gamma}$ can be negative. 
In such a case, diffusioosmotic flow generates positive gauge pressures that raise the rigid plate above the initial gap height to a stable steady state, thus preventing fluid--structure instability.

One interesting extension of the present work, which accounts only for the temporal dynamics of the gap height $h(t)$, is to study the viscous--elastic interaction and the spatiotemporal evolution of the deformable walls of a soft microchannel driven by diffusioosmotic flow.
As a future research direction, it would also be interesting to investigate the coupled dynamics of particle diffusiophoresis and diffusioosmotic flow in soft microfluidic channels, exploring the interplay between diffusiophoresis, diffusioosmosis, and the elasticity of confining boundaries. Diffusioosmotic flow is also encountered as a driving mechanism in nanochannels, where the thin-double-layer assumption may no longer be valid due to comparable Debye length and channel height~\cite{lee2014osmotic}. In such channels, diffusioosmotic flow can generate pressures large enough to deform even relatively rigid walls (e.g., glass covers), making the resulting fluid--structure interaction an interesting direction for further investigation.
We believe that the present study establishes a theoretical foundation for understanding soft microfluidic configurations driven by diffusioosmotic flow, providing key insights for the control and design of fluid--structure interactions and instabilities.

\begin{acknowledgments}
N.M.\ and E.B.\ acknowledge the support by grant no.~1942/23 from the Israel Science Foundation (ISF). E.B.\ acknowledges the support from the Israeli Council for Higher Education Yigal Alon Fellowship.
\end{acknowledgments}

\appendix

\section{Details of a numerical method used in the theoretical model}\label{AppA}

To solve numerically the gap evolution equation (\ref{ND_full_governing_eq}) coupled with the advection--diffusion equation (\ref{adv_diff_with final}) subject to the boundary conditions (\ref{BCs on c ND}) and the initial conditions (\ref{ICs on c and h ND}), we first discretize the spatial derivatives in Eq. (\ref{adv_diff_with final}) using a second-order central difference approximation with uniform grid spacing $\Delta x$. The typical value of the grid size are $\Delta x=0.002$. This discretization leads to a series of ordinary differential equations for the evolution of $c_i(t)=c(x_i,t)$. We then integrate forward in time the resulting set of ordinary differential equations for the evolution of $[h(t),c_i(t)]$ using MATLAB's routine \texttt{ode15s}.
As shown in Figs.~\ref{F3}(b) and~\ref{F8}(b), since the top plate approaches the bottom surface exponentially slowly, we terminate our numerical integration once the gap height $h$ falls below $10^{-2}$.

\section{Details of finite-element numerical simulations}\label{AppB}

\begin{table}
  \begin{center}
  \begin{tabular}{lccc}
Physical property  &   Notation & Value & Units \\
\hline
Initial gap height  & $\tilde{h}_i$ & 20 & $\mu$m\\
Initial concentration  & $\tilde{c}_L$ & 1 & mol m$^{-3}$\\
Length of the rigid top plate  & $\tilde{\ell}$ &  1 & mm \\
Thickness of rigid top plate & $\tilde{b}$ & 10 & $\mu$m \\
Density of rigid top plate & $\tilde{\rho}_s$ & 965 & kg m$^{-3}$ \\
Spring stiffness (per unit depth) & $\tilde{k}$ & 0.125 & Pa\\
Density of fluid  & $\tilde{\rho}$ & $10^3$ & kg m$^{-3}$\\
Viscosity of fluid  & $\tilde{\mu}$ & $10^{-3}$ & Pa s\\

Permittivity of fluid &  $\tilde{\varepsilon}$ &  $7.08\times10^{-10}$ &  $\mathrm{F\,m^{-1}}$\\

Solute diffusivity  & $\tilde{D}$ & $10^{-9}$ & m$^2$ s$^{-1}$ \\
Pressure at the edges  & $\tilde{p}_a$ & 0 & Pa \\
Concentration at the left edge  & $\tilde{c}_L$ & 1 & mol m$^{-3}$ \\
Concentration at the right edge  & $\tilde{c}_R$ & $1-2$ & mol m$^{-3}$ \\
Diffusioosmotic mobility & $\tilde{\Gamma}$ & $10^3$ & $\mu$m$^2$\,s$^{-1}$\\
Characteristic diffusioosmotic slip velocity & $\tilde{u}_c=\tilde{\Gamma}/\tilde{\ell}$ & $1$ & $\mu$m\,s$^{-1}$\\
Characteristic pressure & $\tilde{p}_c = \tilde{\mu}\tilde{\Gamma}/\tilde{h}_i^2$ & $2.5\times10^{-3}$ & Pa\\
Characteristic viscous--elastic time scale  & $\tilde{t}_c = (\tilde{\ell}/\tilde{h}_i)^3(\tilde{\mu}/\tilde{k})$ & $10^3$ & s \\
Axial diffusive time scale & $\tilde{\ell}^2/\tilde{D}$ & $10^{3}$ & s \\
Advection time scale & $\tilde{\ell}/\tilde{u}_c$ & $10^3$ & s \\
      \end{tabular}
  \caption{Representative values of physical parameters used in the two-dimensional finite-element numerical simulations of fluid--structure interaction and fluid--structure instability in microfluidic configurations driven by diffusioosmotic flow.}
  \label{T1}
  \end{center}
\end{table}

In this appendix we provide the details of finite-element numerical simulations.
We perform two-dimensional finite-element simulations with the commercial software
COMSOL Multiphysics (version 6.2, COMSOL AB, Stockholm, Sweden). 

To capture the coupled dynamic behavior of the system, we employ three different modules in the simulations. The first module is the~\textsc{Solid Mechanics module}, which we use to describe the dynamics of the plate. The plate is modeled as a rigid body constrained to vertical motion without rotation and is attached to a spring foundation.
The second module is the~\textsc{Laminar Flow module}, which we employ to describe the fluid mechanics between a rigid bottom surface and a rigid top plate. 
The fluid motion is governed by the Navier--Stokes equations, subject to the boundary conditions on the velocity at the bottom surface and top plate, and fixed pressure at the left and right edges, as given by Eqs.~(\ref{BC_DOS})--(\ref{BC_pressure}).

The~\textsc{Solid Mechanics module} and~\textsc{Laminar Flow module} are fully coupled by the Fluid--Solid Interaction interface, which includes predefined conditions for the interaction at the fluid--solid boundary, $\tilde{z}=\tilde{h}(\tilde{t})$. 
For our case of a rigid top plate attached to a spring foundation, the boundary conditions at the fluid--solid boundary are the kinematic boundary condition, Eq. (\ref{BC_kinamatic_plate}), and the governing equation for the plate motion, Eq.~(\ref{force_eq}), where the fluidic force $\tilde{F}_f$ is given by $\tilde{F}_f=\hat{\boldsymbol{z}} \cdot\int_{0}^{\tilde{\ell}} \hat{\boldsymbol{n}} \cdot \tilde{\boldsymbol{\sigma}}  d\tilde{x}=\int_{0}^{\tilde{\ell}} [\tilde{p}-2\tilde{\mu} (\partial \tilde{w}/\partial \tilde{z} )] d\tilde{x}$. 
Here, $\hat{\boldsymbol{n}}=-\hat{\boldsymbol{z}}$ is the unit normal vector to the fluid--solid boundary and $\tilde{\boldsymbol{\sigma}}=-\tilde{p}\boldsymbol{I}+\tilde{\mu}(\boldsymbol{\nabla}\tilde{\boldsymbol{u}}+(\boldsymbol{\nabla}\tilde{\boldsymbol{u}})^{\mathrm{T}})$ is the stress tensor. 

The third module we use is the~\textsc{Transport of Diluted Species module}. This module describes the spatiotemporal evolution of the solute concentration, which drives diffusioosmotic flow through the slip boundary condition at the bottom surface and, in turn, leads to the fluid--structure interaction. The evolution of the solute concentration is governed by the advection--diffusion equation (\ref{adv-diff}), subject to no-flux boundary conditions at the bottom surface and top plate, and prescribed different concentrations at the left and right edges, as given by Eqs.~(\ref{D_concentartion_bc})--(\ref{D_bc_conc}).

The three modules are fully coupled: the concentration difference between the left and right edges induces fluid motion through the diffusioosmotic slip boundary condition, which exerts a fluidic force on the plate, resulting in its motion and the evolution of the gap height. At the same time, the change in gap height modifies the fluid velocity, the solute concentration distribution, and the magnitude of the applied fluidic force.

We discretize the fluid domain using a rectangular mesh with 500 uniformly distributed elements in the longitudinal dimension and 10 uniformly distributed elements in the transverse dimension.
Due to the time evolution of the gap height, we employ a moving mesh with a fixed bottom surface. The left and right edges of the domain are constrained to prevent deformation in the normal (longitudinal) direction. We use the third-order (cubic) discretization for the velocity field and the second-order (quadratic) discretization for the concentration and pressure distributions.

\begin{table}[b]
  \begin{center}
  \begin{tabular}{lcc}
Non-dimensional number          &   Definition & Value\\
\hline
Aspect ratio  & $\epsilon=\tilde{h}_i/\tilde{\ell}$ & 2$\times10^{-2}$ \\
Reduced Reynolds number & $\epsilon Re = \tilde{\rho}\tilde{u}_c\tilde{h}_i^2/(\tilde{\mu}\tilde{\ell})$ & $4\times10^{-7}$ \\
Womersley number  & $Wo=\tilde{\rho}\tilde{h}_i^2/(\tilde{\mu}\tilde{t}_c)$ & $4\times10^{-7}$  \\
Smallness of rigid plate's inertia & $\tilde{\rho}_s\tilde{b}\tilde{\ell}/(\tilde{k}\tilde{t}_c^2)$ & $7.7\times10^{-11}$ \\
Smallness of dielectric effect (KCL with $\beta_D=0$)& $\beta_D^2\tilde{\varepsilon}\tilde{\varphi}_T^2 \tilde{h}_{i}^2 /(\tilde{\mu}\tilde{\Gamma}\tilde{\ell}^2)$ &  0   \\
Smallness of dielectric effect (NaCL with $\beta_D=-0.208$)& $\beta_D^2\tilde{\varepsilon}\tilde{\varphi}_T^2 \tilde{h}_{i}^2 /(\tilde{\mu}\tilde{\Gamma}\tilde{\ell}^2)$ &  $7.7\times10^{-6}$   \\
Elasto-diffusioosmotic number & $\mathcal{F}_{\rm DOF}=81\tilde{\mu}\tilde{u}_c\tilde{\ell}^2/(2\tilde{k}\tilde{h}_i^3)$ & 40.5 \\
Elasto-diffusion number  & $\alpha =\tilde{\mu}\tilde{D}\tilde{\ell}/(\tilde{k}\tilde{h}_i^3)$ & 1 \\
P\'{e}clet number & $Pe=\tilde{u}_c\tilde{\ell}/\tilde{D}$ & 1  \\
Non-dimensional right-edge concentration & $c_R=\tilde{c}_R/\tilde{c}_L$ & $1-2$ \\
      \end{tabular}
  \caption{Representative values of non-dimensional numbers corresponding to the physical parameters in Table~\ref{T1}, indicating that the assumptions of the theoretical model are well satisfied. The values of $\beta_D = (\tilde{D}_{+}-\tilde{D}_{-})/(\tilde{D}_{+}+\tilde{D}_{-})$ were calculated using $\tilde{D}_{+}=1.33\times 10^{-9}$ m$^2$\,s$^{-1}$ for $\text{Na}^{+}$, $\tilde{D}_{+}=2.03\times 10^{-9}$ m$^2$\,s$^{-1}$ for $\text{K}^{+}$, and $\tilde{D}_{-}=2.03\times 10^{-9}$ m$^2$\,s$^{-1}$ for $\text{Cl}^{-}$~\cite{vanysek1993ionic,shin2016size}.}
  \label{T2}
  \end{center}
\end{table}

We obtained all finite-element simulation results presented in this study through dynamic simulations.
For the unstable cases, we stopped the simulations when the gap height fell below $0.2~\mu$m, corresponding to $1\%$ of the initial gap height. For the stable cases, we stopped the simulations after $10^7$ s, once the system reached a steady state with negligible plate motion.

To assess the grid sensitivity, we performed tests by considering seven different
mesh resolutions ranging from $100 \times 2$ to $700 \times 14$ elements across 16 values of the right-edge concentration $\tilde{c}_R$
between~1.561 mol m$^{-3}$  and 1.564 mol m$^{-3}$, encompassing the critical threshold for the onset of instability. We established grid independence using our selected mesh, with the maximum relative error in the steady-state gap height remaining below $0.15\%$ compared to the finest mesh across all tested concentration values.

We summarize in Table~\ref{T1} the representative values of physical and geometrical parameters used in the finite-element simulations. Table~\ref{T2} summarizes the corresponding non-dimensional numbers, showing that our theoretical model's assumptions are well satisfied.

\bibliography{literature}

\begin{thebibliography}{70}%
\makeatletter
\providecommand \@ifxundefined [1]{%
 \@ifx{#1\undefined}
}%
\providecommand \@ifnum [1]{%
 \ifnum #1\expandafter \@firstoftwo
 \else \expandafter \@secondoftwo
 \fi
}%
\providecommand \@ifx [1]{%
 \ifx #1\expandafter \@firstoftwo
 \else \expandafter \@secondoftwo
 \fi
}%
\providecommand \natexlab [1]{#1}%
\providecommand \enquote  [1]{``#1''}%
\providecommand \bibnamefont  [1]{#1}%
\providecommand \bibfnamefont [1]{#1}%
\providecommand \citenamefont [1]{#1}%
\providecommand \href@noop [0]{\@secondoftwo}%
\providecommand \href [0]{\begingroup \@sanitize@url \@href}%
\providecommand \@href[1]{\@@startlink{#1}\@@href}%
\providecommand \@@href[1]{\endgroup#1\@@endlink}%
\providecommand \@sanitize@url [0]{\catcode `\\12\catcode `\$12\catcode `\&12\catcode `\#12\catcode `\^12\catcode `\_12\catcode `\%12\relax}%
\providecommand \@@startlink[1]{}%
\providecommand \@@endlink[0]{}%
\providecommand \url  [0]{\begingroup\@sanitize@url \@url }%
\providecommand \@url [1]{\endgroup\@href {#1}{\urlprefix }}%
\providecommand \urlprefix  [0]{URL }%
\providecommand \Eprint [0]{\href }%
\providecommand \doibase [0]{https://doi.org/}%
\providecommand \selectlanguage [0]{\@gobble}%
\providecommand \bibinfo  [0]{\@secondoftwo}%
\providecommand \bibfield  [0]{\@secondoftwo}%
\providecommand \translation [1]{[#1]}%
\providecommand \BibitemOpen [0]{}%
\providecommand \bibitemStop [0]{}%
\providecommand \bibitemNoStop [0]{.\EOS\space}%
\providecommand \EOS [0]{\spacefactor3000\relax}%
\providecommand \BibitemShut  [1]{\csname bibitem#1\endcsname}%
\let\auto@bib@innerbib\@empty
\bibitem [{\citenamefont {Prieve}\ \emph {et~al.}(1984)\citenamefont {Prieve}, \citenamefont {Anderson}, \citenamefont {Ebel},\ and\ \citenamefont {Lowell}}]{prieve1984motion}%
  \BibitemOpen
  \bibfield  {author} {\bibinfo {author} {\bibfnamefont {D.~C.}\ \bibnamefont {Prieve}}, \bibinfo {author} {\bibfnamefont {J.~L.}\ \bibnamefont {Anderson}}, \bibinfo {author} {\bibfnamefont {J.~P.}\ \bibnamefont {Ebel}},\ and\ \bibinfo {author} {\bibfnamefont {M.~E.}\ \bibnamefont {Lowell}},\ }\bibfield  {title} {\bibinfo {title} {Motion of a particle generated by chemical gradients. {Part 2. Electrolytes}},\ }\href@noop {} {\bibfield  {journal} {\bibinfo  {journal} {J. Fluid Mech.}\ }\textbf {\bibinfo {volume} {148}},\ \bibinfo {pages} {247} (\bibinfo {year} {1984})}\BibitemShut {NoStop}%
\bibitem [{\citenamefont {Anderson}(1989)}]{anderson1989colloid}%
  \BibitemOpen
  \bibfield  {author} {\bibinfo {author} {\bibfnamefont {J.~L.}\ \bibnamefont {Anderson}},\ }\bibfield  {title} {\bibinfo {title} {Colloid transport by interfacial forces},\ }\href@noop {} {\bibfield  {journal} {\bibinfo  {journal} {Annu. Rev. Fluid Mech.}\ }\textbf {\bibinfo {volume} {21}},\ \bibinfo {pages} {61} (\bibinfo {year} {1989})}\BibitemShut {NoStop}%
\bibitem [{\citenamefont {Shim}(2022)}]{shim2022diffusiophoresis}%
  \BibitemOpen
  \bibfield  {author} {\bibinfo {author} {\bibfnamefont {S.}~\bibnamefont {Shim}},\ }\bibfield  {title} {\bibinfo {title} {Diffusiophoresis, diffusioosmosis, and microfluidics: surface-flow-driven phenomena in the presence of flow},\ }\href@noop {} {\bibfield  {journal} {\bibinfo  {journal} {Chem. Rev.}\ }\textbf {\bibinfo {volume} {122}},\ \bibinfo {pages} {6986} (\bibinfo {year} {2022})}\BibitemShut {NoStop}%
\bibitem [{\citenamefont {Ault}\ and\ \citenamefont {Shin}(2025)}]{ault2024physicochemical}%
  \BibitemOpen
  \bibfield  {author} {\bibinfo {author} {\bibfnamefont {J.~T.}\ \bibnamefont {Ault}}\ and\ \bibinfo {author} {\bibfnamefont {S.}~\bibnamefont {Shin}},\ }\bibfield  {title} {\bibinfo {title} {Physicochemical hydrodynamics of particle diffusiophoresis driven by chemical gradients},\ }\href@noop {} {\bibfield  {journal} {\bibinfo  {journal} {Annu. Rev. Fluid Mech.}\ }\textbf {\bibinfo {volume} {57}},\ \bibinfo {pages} {227} (\bibinfo {year} {2025})}\BibitemShut {NoStop}%
\bibitem [{\citenamefont {Shi}\ \emph {et~al.}(2025)\citenamefont {Shi}, \citenamefont {Fattah},\ and\ \citenamefont {Squires}}]{shi2025generalized}%
  \BibitemOpen
  \bibfield  {author} {\bibinfo {author} {\bibfnamefont {N.}~\bibnamefont {Shi}}, \bibinfo {author} {\bibfnamefont {A.}~\bibnamefont {Fattah}},\ and\ \bibinfo {author} {\bibfnamefont {T.~M.}\ \bibnamefont {Squires}},\ }\bibfield  {title} {\bibinfo {title} {Generalized, conceptually unified theory of linear phoretic drift and osmotic slip},\ }\href@noop {} {\bibfield  {journal} {\bibinfo  {journal} {Phys. Rev. Fluids}\ }\textbf {\bibinfo {volume} {10}},\ \bibinfo {pages} {043701} (\bibinfo {year} {2025})}\BibitemShut {NoStop}%
\bibitem [{\citenamefont {Derjaguin}\ \emph {et~al.}(1961)\citenamefont {Derjaguin}, \citenamefont {Dukhin},\ and\ \citenamefont {Korotkova}}]{derjaguin1961diffusiophoresis}%
  \BibitemOpen
  \bibfield  {author} {\bibinfo {author} {\bibfnamefont {B.~V.}\ \bibnamefont {Derjaguin}}, \bibinfo {author} {\bibfnamefont {S.~S.}\ \bibnamefont {Dukhin}},\ and\ \bibinfo {author} {\bibfnamefont {A.~A.}\ \bibnamefont {Korotkova}},\ }\bibfield  {title} {\bibinfo {title} {Diffusiophoresis in electrolyte solutions and its role in the mechanism of film formation from rubber latexes by the method of ionic deposition},\ }\href@noop {} {\bibfield  {journal} {\bibinfo  {journal} {Kolloidn. Zh.}\ }\textbf {\bibinfo {volume} {23}},\ \bibinfo {pages} {53} (\bibinfo {year} {1961})}\BibitemShut {NoStop}%
\bibitem [{\citenamefont {Derjaguin}\ \emph {et~al.}(1993)\citenamefont {Derjaguin}, \citenamefont {Dukhin},\ and\ \citenamefont {Korotkova}}]{derjaguin1993diffusiophoresis}%
  \BibitemOpen
  \bibfield  {author} {\bibinfo {author} {\bibfnamefont {B.~V.}\ \bibnamefont {Derjaguin}}, \bibinfo {author} {\bibfnamefont {S.~S.}\ \bibnamefont {Dukhin}},\ and\ \bibinfo {author} {\bibfnamefont {A.~A.}\ \bibnamefont {Korotkova}},\ }\bibfield  {title} {\bibinfo {title} {{Diffusiophoresis in electrolyte solutions and its role in the Mechanism of the formation of films from caoutchouc latexes by the ionic deposition method}},\ }\href@noop {} {\bibfield  {journal} {\bibinfo  {journal} {Prog. Surf. Sci.}\ }\textbf {\bibinfo {volume} {43}},\ \bibinfo {pages} {153} (\bibinfo {year} {1993})}\BibitemShut {NoStop}%
\bibitem [{\citenamefont {Lee}\ \emph {et~al.}(2014)\citenamefont {Lee}, \citenamefont {Cottin-Bizonne}, \citenamefont {Biance}, \citenamefont {Joseph}, \citenamefont {Bocquet},\ and\ \citenamefont {Ybert}}]{lee2014osmotic}%
  \BibitemOpen
  \bibfield  {author} {\bibinfo {author} {\bibfnamefont {C.}~\bibnamefont {Lee}}, \bibinfo {author} {\bibfnamefont {C.}~\bibnamefont {Cottin-Bizonne}}, \bibinfo {author} {\bibfnamefont {A.-L.}\ \bibnamefont {Biance}}, \bibinfo {author} {\bibfnamefont {P.}~\bibnamefont {Joseph}}, \bibinfo {author} {\bibfnamefont {L.}~\bibnamefont {Bocquet}},\ and\ \bibinfo {author} {\bibfnamefont {C.}~\bibnamefont {Ybert}},\ }\bibfield  {title} {\bibinfo {title} {Osmotic flow through fully permeable nanochannels},\ }\href@noop {} {\bibfield  {journal} {\bibinfo  {journal} {Phys. Rev. Lett.}\ }\textbf {\bibinfo {volume} {112}},\ \bibinfo {pages} {244501} (\bibinfo {year} {2014})}\BibitemShut {NoStop}%
\bibitem [{\citenamefont {Liu}\ and\ \citenamefont {Pahlavan}(2025)}]{liu2025diffusioosmotic}%
  \BibitemOpen
  \bibfield  {author} {\bibinfo {author} {\bibfnamefont {H.}~\bibnamefont {Liu}}\ and\ \bibinfo {author} {\bibfnamefont {A.~A.}\ \bibnamefont {Pahlavan}},\ }\bibfield  {title} {\bibinfo {title} {{Diffusioosmotic reversal of colloidal focusing direction in a microfluidic T-junction}},\ }\href@noop {} {\bibfield  {journal} {\bibinfo  {journal} {Phys. Rev. Lett.}\ }\textbf {\bibinfo {volume} {134}},\ \bibinfo {pages} {098201} (\bibinfo {year} {2025})}\BibitemShut {NoStop}%
\bibitem [{\citenamefont {Stone}\ \emph {et~al.}(2004)\citenamefont {Stone}, \citenamefont {Stroock},\ and\ \citenamefont {Ajdari}}]{stone2004engineering}%
  \BibitemOpen
  \bibfield  {author} {\bibinfo {author} {\bibfnamefont {H.~A.}\ \bibnamefont {Stone}}, \bibinfo {author} {\bibfnamefont {A.~D.}\ \bibnamefont {Stroock}},\ and\ \bibinfo {author} {\bibfnamefont {A.}~\bibnamefont {Ajdari}},\ }\bibfield  {title} {\bibinfo {title} {Engineering flows in small devices: microfluidics toward a lab-on-a-chip},\ }\href@noop {} {\bibfield  {journal} {\bibinfo  {journal} {Annu. Rev. Fluid Mech.}\ }\textbf {\bibinfo {volume} {36}},\ \bibinfo {pages} {381} (\bibinfo {year} {2004})}\BibitemShut {NoStop}%
\bibitem [{\citenamefont {Squires}\ and\ \citenamefont {Quake}(2005)}]{squires2005microfluidics}%
  \BibitemOpen
  \bibfield  {author} {\bibinfo {author} {\bibfnamefont {T.~M.}\ \bibnamefont {Squires}}\ and\ \bibinfo {author} {\bibfnamefont {S.~R.}\ \bibnamefont {Quake}},\ }\bibfield  {title} {\bibinfo {title} {{Microfluidics: Fluid physics at the nanoliter scale}},\ }\href@noop {} {\bibfield  {journal} {\bibinfo  {journal} {Rev. Mod. Phys.}\ }\textbf {\bibinfo {volume} {77}},\ \bibinfo {pages} {977} (\bibinfo {year} {2005})}\BibitemShut {NoStop}%
\bibitem [{\citenamefont {Paratore}\ \emph {et~al.}(2022)\citenamefont {Paratore}, \citenamefont {Bacheva}, \citenamefont {Bercovici},\ and\ \citenamefont {Kaigala}}]{paratore2022reconfigurable}%
  \BibitemOpen
  \bibfield  {author} {\bibinfo {author} {\bibfnamefont {F.}~\bibnamefont {Paratore}}, \bibinfo {author} {\bibfnamefont {V.}~\bibnamefont {Bacheva}}, \bibinfo {author} {\bibfnamefont {M.}~\bibnamefont {Bercovici}},\ and\ \bibinfo {author} {\bibfnamefont {G.~V.}\ \bibnamefont {Kaigala}},\ }\bibfield  {title} {\bibinfo {title} {Reconfigurable microfluidics},\ }\href@noop {} {\bibfield  {journal} {\bibinfo  {journal} {Nat. Rev. Chem.}\ }\textbf {\bibinfo {volume} {6}},\ \bibinfo {pages} {70} (\bibinfo {year} {2022})}\BibitemShut {NoStop}%
\bibitem [{\citenamefont {Ault}\ \emph {et~al.}(2019)\citenamefont {Ault}, \citenamefont {Shin},\ and\ \citenamefont {Stone}}]{ault2019characterization}%
  \BibitemOpen
  \bibfield  {author} {\bibinfo {author} {\bibfnamefont {J.~T.}\ \bibnamefont {Ault}}, \bibinfo {author} {\bibfnamefont {S.}~\bibnamefont {Shin}},\ and\ \bibinfo {author} {\bibfnamefont {H.~A.}\ \bibnamefont {Stone}},\ }\bibfield  {title} {\bibinfo {title} {Characterization of surface--solute interactions by diffusioosmosis},\ }\href@noop {} {\bibfield  {journal} {\bibinfo  {journal} {Soft Matt.}\ }\textbf {\bibinfo {volume} {15}},\ \bibinfo {pages} {1582} (\bibinfo {year} {2019})}\BibitemShut {NoStop}%
\bibitem [{\citenamefont {Teng}\ \emph {et~al.}(2023)\citenamefont {Teng}, \citenamefont {Rallabandi},\ and\ \citenamefont {Ault}}]{teng2023diffusioosmotic}%
  \BibitemOpen
  \bibfield  {author} {\bibinfo {author} {\bibfnamefont {J.}~\bibnamefont {Teng}}, \bibinfo {author} {\bibfnamefont {B.}~\bibnamefont {Rallabandi}},\ and\ \bibinfo {author} {\bibfnamefont {J.~T.}\ \bibnamefont {Ault}},\ }\bibfield  {title} {\bibinfo {title} {Diffusioosmotic dispersion of solute in a long narrow channel},\ }\href@noop {} {\bibfield  {journal} {\bibinfo  {journal} {J. Fluid Mech.}\ }\textbf {\bibinfo {volume} {977}},\ \bibinfo {pages} {A5} (\bibinfo {year} {2023})}\BibitemShut {NoStop}%
\bibitem [{\citenamefont {Ault}\ \emph {et~al.}(2017)\citenamefont {Ault}, \citenamefont {Warren}, \citenamefont {Shin},\ and\ \citenamefont {Stone}}]{ault2017diffusiophoresis}%
  \BibitemOpen
  \bibfield  {author} {\bibinfo {author} {\bibfnamefont {J.~T.}\ \bibnamefont {Ault}}, \bibinfo {author} {\bibfnamefont {P.~B.}\ \bibnamefont {Warren}}, \bibinfo {author} {\bibfnamefont {S.}~\bibnamefont {Shin}},\ and\ \bibinfo {author} {\bibfnamefont {H.~A.}\ \bibnamefont {Stone}},\ }\bibfield  {title} {\bibinfo {title} {Diffusiophoresis in one-dimensional solute gradients},\ }\href@noop {} {\bibfield  {journal} {\bibinfo  {journal} {Soft Matt.}\ }\textbf {\bibinfo {volume} {13}},\ \bibinfo {pages} {9015} (\bibinfo {year} {2017})}\BibitemShut {NoStop}%
\bibitem [{\citenamefont {Gupta}\ \emph {et~al.}(2020)\citenamefont {Gupta}, \citenamefont {Shim},\ and\ \citenamefont {Stone}}]{gupta2020diffusiophoresis}%
  \BibitemOpen
  \bibfield  {author} {\bibinfo {author} {\bibfnamefont {A.}~\bibnamefont {Gupta}}, \bibinfo {author} {\bibfnamefont {S.}~\bibnamefont {Shim}},\ and\ \bibinfo {author} {\bibfnamefont {H.~A.}\ \bibnamefont {Stone}},\ }\bibfield  {title} {\bibinfo {title} {Diffusiophoresis: from dilute to concentrated electrolytes},\ }\href@noop {} {\bibfield  {journal} {\bibinfo  {journal} {Soft Matt.}\ }\textbf {\bibinfo {volume} {16}},\ \bibinfo {pages} {6975} (\bibinfo {year} {2020})}\BibitemShut {NoStop}%
\bibitem [{\citenamefont {Singh}\ \emph {et~al.}(2020)\citenamefont {Singh}, \citenamefont {Vladisavljevi{\'c}}, \citenamefont {Nadal}, \citenamefont {Cottin-Bizonne}, \citenamefont {Pirat},\ and\ \citenamefont {Bolognesi}}]{singh2020reversible}%
  \BibitemOpen
  \bibfield  {author} {\bibinfo {author} {\bibfnamefont {N.}~\bibnamefont {Singh}}, \bibinfo {author} {\bibfnamefont {G.~T.}\ \bibnamefont {Vladisavljevi{\'c}}}, \bibinfo {author} {\bibfnamefont {F.}~\bibnamefont {Nadal}}, \bibinfo {author} {\bibfnamefont {C.}~\bibnamefont {Cottin-Bizonne}}, \bibinfo {author} {\bibfnamefont {C.}~\bibnamefont {Pirat}},\ and\ \bibinfo {author} {\bibfnamefont {G.}~\bibnamefont {Bolognesi}},\ }\bibfield  {title} {\bibinfo {title} {Reversible trapping of colloids in microgrooved channels via diffusiophoresis under steady-state solute gradients},\ }\href@noop {} {\bibfield  {journal} {\bibinfo  {journal} {Phys. Rev. Lett.}\ }\textbf {\bibinfo {volume} {125}},\ \bibinfo {pages} {248002} (\bibinfo {year} {2020})}\BibitemShut {NoStop}%
\bibitem [{\citenamefont {Singh}\ \emph {et~al.}(2022)\citenamefont {Singh}, \citenamefont {Vladisavljevi{\'c}}, \citenamefont {Nadal}, \citenamefont {Cottin-Bizonne}, \citenamefont {Pirat},\ and\ \citenamefont {Bolognesi}}]{singh2022enhanced}%
  \BibitemOpen
  \bibfield  {author} {\bibinfo {author} {\bibfnamefont {N.}~\bibnamefont {Singh}}, \bibinfo {author} {\bibfnamefont {G.~T.}\ \bibnamefont {Vladisavljevi{\'c}}}, \bibinfo {author} {\bibfnamefont {F.}~\bibnamefont {Nadal}}, \bibinfo {author} {\bibfnamefont {C.}~\bibnamefont {Cottin-Bizonne}}, \bibinfo {author} {\bibfnamefont {C.}~\bibnamefont {Pirat}},\ and\ \bibinfo {author} {\bibfnamefont {G.}~\bibnamefont {Bolognesi}},\ }\bibfield  {title} {\bibinfo {title} {Enhanced accumulation of colloidal particles in microgrooved channels via diffusiophoresis and steady-state electrolyte flows},\ }\href@noop {} {\bibfield  {journal} {\bibinfo  {journal} {Langmuir}\ }\textbf {\bibinfo {volume} {38}},\ \bibinfo {pages} {14053} (\bibinfo {year} {2022})}\BibitemShut {NoStop}%
\bibitem [{\citenamefont {Migacz}\ \emph {et~al.}(2024)\citenamefont {Migacz}, \citenamefont {Castleberry},\ and\ \citenamefont {Ault}}]{migacz2024enhanced}%
  \BibitemOpen
  \bibfield  {author} {\bibinfo {author} {\bibfnamefont {R.~E.}\ \bibnamefont {Migacz}}, \bibinfo {author} {\bibfnamefont {M.}~\bibnamefont {Castleberry}},\ and\ \bibinfo {author} {\bibfnamefont {J.~T.}\ \bibnamefont {Ault}},\ }\bibfield  {title} {\bibinfo {title} {Enhanced diffusiophoresis in dead-end pores with time-dependent boundary solute concentration},\ }\href@noop {} {\bibfield  {journal} {\bibinfo  {journal} {Phys. Rev. Fluids}\ }\textbf {\bibinfo {volume} {9}},\ \bibinfo {pages} {044203} (\bibinfo {year} {2024})}\BibitemShut {NoStop}%
\bibitem [{\citenamefont {Kar}\ \emph {et~al.}(2015)\citenamefont {Kar}, \citenamefont {Chiang}, \citenamefont {Ortiz~Rivera}, \citenamefont {Sen},\ and\ \citenamefont {Velegol}}]{kar2015enhanced}%
  \BibitemOpen
  \bibfield  {author} {\bibinfo {author} {\bibfnamefont {A.}~\bibnamefont {Kar}}, \bibinfo {author} {\bibfnamefont {T.}~\bibnamefont {Chiang}}, \bibinfo {author} {\bibfnamefont {I.}~\bibnamefont {Ortiz~Rivera}}, \bibinfo {author} {\bibfnamefont {A.}~\bibnamefont {Sen}},\ and\ \bibinfo {author} {\bibfnamefont {D.}~\bibnamefont {Velegol}},\ }\bibfield  {title} {\bibinfo {title} {Enhanced transport into and out of dead-end pores},\ }\href@noop {} {\bibfield  {journal} {\bibinfo  {journal} {ACS Nano}\ }\textbf {\bibinfo {volume} {9}},\ \bibinfo {pages} {746} (\bibinfo {year} {2015})}\BibitemShut {NoStop}%
\bibitem [{\citenamefont {Shin}\ \emph {et~al.}(2017)\citenamefont {Shin}, \citenamefont {Ault}, \citenamefont {Warren},\ and\ \citenamefont {Stone}}]{shin2017accumulation}%
  \BibitemOpen
  \bibfield  {author} {\bibinfo {author} {\bibfnamefont {S.}~\bibnamefont {Shin}}, \bibinfo {author} {\bibfnamefont {J.~T.}\ \bibnamefont {Ault}}, \bibinfo {author} {\bibfnamefont {P.~B.}\ \bibnamefont {Warren}},\ and\ \bibinfo {author} {\bibfnamefont {H.~A.}\ \bibnamefont {Stone}},\ }\bibfield  {title} {\bibinfo {title} {Accumulation of colloidal particles in flow junctions induced by fluid flow and diffusiophoresis},\ }\href@noop {} {\bibfield  {journal} {\bibinfo  {journal} {Phys. Rev. X}\ }\textbf {\bibinfo {volume} {7}},\ \bibinfo {pages} {041038} (\bibinfo {year} {2017})}\BibitemShut {NoStop}%
\bibitem [{\citenamefont {Shin}\ \emph {et~al.}(2016)\citenamefont {Shin}, \citenamefont {Um}, \citenamefont {Sabass}, \citenamefont {Ault}, \citenamefont {Rahimi}, \citenamefont {Warren},\ and\ \citenamefont {Stone}}]{shin2016size}%
  \BibitemOpen
  \bibfield  {author} {\bibinfo {author} {\bibfnamefont {S.}~\bibnamefont {Shin}}, \bibinfo {author} {\bibfnamefont {E.}~\bibnamefont {Um}}, \bibinfo {author} {\bibfnamefont {B.}~\bibnamefont {Sabass}}, \bibinfo {author} {\bibfnamefont {J.~T.}\ \bibnamefont {Ault}}, \bibinfo {author} {\bibfnamefont {M.}~\bibnamefont {Rahimi}}, \bibinfo {author} {\bibfnamefont {P.~B.}\ \bibnamefont {Warren}},\ and\ \bibinfo {author} {\bibfnamefont {H.~A.}\ \bibnamefont {Stone}},\ }\bibfield  {title} {\bibinfo {title} {Size-dependent control of colloid transport via solute gradients in dead-end channels},\ }\href@noop {} {\bibfield  {journal} {\bibinfo  {journal} {Proc. Natl Acad. Sci. USA}\ }\textbf {\bibinfo {volume} {113}},\ \bibinfo {pages} {257} (\bibinfo {year} {2016})}\BibitemShut {NoStop}%
\bibitem [{\citenamefont {Ault}\ \emph {et~al.}(2018)\citenamefont {Ault}, \citenamefont {Shin},\ and\ \citenamefont {Stone}}]{ault2018diffusiophoresis}%
  \BibitemOpen
  \bibfield  {author} {\bibinfo {author} {\bibfnamefont {J.~T.}\ \bibnamefont {Ault}}, \bibinfo {author} {\bibfnamefont {S.}~\bibnamefont {Shin}},\ and\ \bibinfo {author} {\bibfnamefont {H.~A.}\ \bibnamefont {Stone}},\ }\bibfield  {title} {\bibinfo {title} {Diffusiophoresis in narrow channel flows},\ }\href@noop {} {\bibfield  {journal} {\bibinfo  {journal} {J. Fluid Mech.}\ }\textbf {\bibinfo {volume} {854}},\ \bibinfo {pages} {420} (\bibinfo {year} {2018})}\BibitemShut {NoStop}%
\bibitem [{\citenamefont {Ha}\ \emph {et~al.}(2019)\citenamefont {Ha}, \citenamefont {Seo}, \citenamefont {Lee},\ and\ \citenamefont {Kim}}]{ha2019dynamic}%
  \BibitemOpen
  \bibfield  {author} {\bibinfo {author} {\bibfnamefont {D.}~\bibnamefont {Ha}}, \bibinfo {author} {\bibfnamefont {S.}~\bibnamefont {Seo}}, \bibinfo {author} {\bibfnamefont {K.}~\bibnamefont {Lee}},\ and\ \bibinfo {author} {\bibfnamefont {T.}~\bibnamefont {Kim}},\ }\bibfield  {title} {\bibinfo {title} {Dynamic transport control of colloidal particles by repeatable active switching of solute gradients},\ }\href@noop {} {\bibfield  {journal} {\bibinfo  {journal} {ACS Nano}\ }\textbf {\bibinfo {volume} {13}},\ \bibinfo {pages} {12939} (\bibinfo {year} {2019})}\BibitemShut {NoStop}%
\bibitem [{\citenamefont {Alessio}\ \emph {et~al.}(2021)\citenamefont {Alessio}, \citenamefont {Shim}, \citenamefont {Mintah}, \citenamefont {Gupta},\ and\ \citenamefont {Stone}}]{alessio2021diffusiophoresis}%
  \BibitemOpen
  \bibfield  {author} {\bibinfo {author} {\bibfnamefont {B.~M.}\ \bibnamefont {Alessio}}, \bibinfo {author} {\bibfnamefont {S.}~\bibnamefont {Shim}}, \bibinfo {author} {\bibfnamefont {E.}~\bibnamefont {Mintah}}, \bibinfo {author} {\bibfnamefont {A.}~\bibnamefont {Gupta}},\ and\ \bibinfo {author} {\bibfnamefont {H.~A.}\ \bibnamefont {Stone}},\ }\bibfield  {title} {\bibinfo {title} {{Diffusiophoresis and diffusioosmosis in tandem: Two-dimensional particle motion in the presence of multiple electrolytes}},\ }\href@noop {} {\bibfield  {journal} {\bibinfo  {journal} {Phys. Rev. Fluids}\ }\textbf {\bibinfo {volume} {6}},\ \bibinfo {pages} {054201} (\bibinfo {year} {2021})}\BibitemShut {NoStop}%
\bibitem [{\citenamefont {Alessio}\ \emph {et~al.}(2022)\citenamefont {Alessio}, \citenamefont {Shim}, \citenamefont {Gupta},\ and\ \citenamefont {Stone}}]{alessio2022diffusioosmosis}%
  \BibitemOpen
  \bibfield  {author} {\bibinfo {author} {\bibfnamefont {B.~M.}\ \bibnamefont {Alessio}}, \bibinfo {author} {\bibfnamefont {S.}~\bibnamefont {Shim}}, \bibinfo {author} {\bibfnamefont {A.}~\bibnamefont {Gupta}},\ and\ \bibinfo {author} {\bibfnamefont {H.~A.}\ \bibnamefont {Stone}},\ }\bibfield  {title} {\bibinfo {title} {{Diffusioosmosis-driven dispersion of colloids: a Taylor dispersion analysis with experimental validation}},\ }\href@noop {} {\bibfield  {journal} {\bibinfo  {journal} {J. Fluid Mech.}\ }\textbf {\bibinfo {volume} {942}},\ \bibinfo {pages} {A23} (\bibinfo {year} {2022})}\BibitemShut {NoStop}%
\bibitem [{\citenamefont {Akdeniz}\ \emph {et~al.}(2023)\citenamefont {Akdeniz}, \citenamefont {Wood},\ and\ \citenamefont {Lammertink}}]{akdeniz2023diffusiophoresis}%
  \BibitemOpen
  \bibfield  {author} {\bibinfo {author} {\bibfnamefont {B.}~\bibnamefont {Akdeniz}}, \bibinfo {author} {\bibfnamefont {J.~A.}\ \bibnamefont {Wood}},\ and\ \bibinfo {author} {\bibfnamefont {R.~G.~H.}\ \bibnamefont {Lammertink}},\ }\bibfield  {title} {\bibinfo {title} {{Diffusiophoresis and diffusio-osmosis into a dead-end channel: Role of the concentration-dependence of zeta potential}},\ }\href@noop {} {\bibfield  {journal} {\bibinfo  {journal} {Langmuir}\ }\textbf {\bibinfo {volume} {39}},\ \bibinfo {pages} {2322} (\bibinfo {year} {2023})}\BibitemShut {NoStop}%
\bibitem [{\citenamefont {Whitesides}(2006)}]{whitesides2006origins}%
  \BibitemOpen
  \bibfield  {author} {\bibinfo {author} {\bibfnamefont {G.~M.}\ \bibnamefont {Whitesides}},\ }\bibfield  {title} {\bibinfo {title} {The origins and the future of microfluidics},\ }\href@noop {} {\bibfield  {journal} {\bibinfo  {journal} {Nature}\ }\textbf {\bibinfo {volume} {442}},\ \bibinfo {pages} {368} (\bibinfo {year} {2006})}\BibitemShut {NoStop}%
\bibitem [{\citenamefont {Christov}(2022)}]{christov2021soft}%
  \BibitemOpen
  \bibfield  {author} {\bibinfo {author} {\bibfnamefont {I.~C.}\ \bibnamefont {Christov}},\ }\bibfield  {title} {\bibinfo {title} {Soft hydraulics: from {N}ewtonian to complex fluid flows through compliant conduits},\ }\href@noop {} {\bibfield  {journal} {\bibinfo  {journal} {J. Phys.: Condens. Matter}\ }\textbf {\bibinfo {volume} {38}},\ \bibinfo {pages} {063001} (\bibinfo {year} {2022})}\BibitemShut {NoStop}%
\bibitem [{\citenamefont {Kashaninejad}\ and\ \citenamefont {Nguyen}(2023)}]{kashaninejad2023microfluidic}%
  \BibitemOpen
  \bibfield  {author} {\bibinfo {author} {\bibfnamefont {N.}~\bibnamefont {Kashaninejad}}\ and\ \bibinfo {author} {\bibfnamefont {N.-T.}\ \bibnamefont {Nguyen}},\ }\bibfield  {title} {\bibinfo {title} {Microfluidic solutions for biofluids handling in on-skin wearable systems},\ }\href@noop {} {\bibfield  {journal} {\bibinfo  {journal} {Lab Chip}\ }\textbf {\bibinfo {volume} {23}},\ \bibinfo {pages} {913} (\bibinfo {year} {2023})}\BibitemShut {NoStop}%
\bibitem [{\citenamefont {Grotberg}\ and\ \citenamefont {Jensen}(2004)}]{grotberg2004biofluid}%
  \BibitemOpen
  \bibfield  {author} {\bibinfo {author} {\bibfnamefont {J.~B.}\ \bibnamefont {Grotberg}}\ and\ \bibinfo {author} {\bibfnamefont {O.~E.}\ \bibnamefont {Jensen}},\ }\bibfield  {title} {\bibinfo {title} {Biofluid mechanics in flexible tubes},\ }\href@noop {} {\bibfield  {journal} {\bibinfo  {journal} {Annu. Rev. Fluid Mech.}\ }\textbf {\bibinfo {volume} {36}},\ \bibinfo {pages} {121} (\bibinfo {year} {2004})}\BibitemShut {NoStop}%
\bibitem [{\citenamefont {Bhatia}\ and\ \citenamefont {Ingber}(2014)}]{bhatia2014microfluidic}%
  \BibitemOpen
  \bibfield  {author} {\bibinfo {author} {\bibfnamefont {S.~N.}\ \bibnamefont {Bhatia}}\ and\ \bibinfo {author} {\bibfnamefont {D.~E.}\ \bibnamefont {Ingber}},\ }\bibfield  {title} {\bibinfo {title} {Microfluidic organs-on-chips},\ }\href@noop {} {\bibfield  {journal} {\bibinfo  {journal} {Nat. Biotechnol.}\ }\textbf {\bibinfo {volume} {32}},\ \bibinfo {pages} {760} (\bibinfo {year} {2014})}\BibitemShut {NoStop}%
\bibitem [{\citenamefont {Al-Housseiny}\ and\ \citenamefont {Stone}(2013)}]{al2013controlling}%
  \BibitemOpen
  \bibfield  {author} {\bibinfo {author} {\bibfnamefont {T.~T.}\ \bibnamefont {Al-Housseiny}}\ and\ \bibinfo {author} {\bibfnamefont {H.~A.}\ \bibnamefont {Stone}},\ }\bibfield  {title} {\bibinfo {title} {{Controlling viscous fingering in tapered Hele-Shaw cells}},\ }\href@noop {} {\bibfield  {journal} {\bibinfo  {journal} {Phys. Fluids}\ }\textbf {\bibinfo {volume} {25}} (\bibinfo {year} {2013})}\BibitemShut {NoStop}%
\bibitem [{\citenamefont {Fontana}\ \emph {et~al.}(2024)\citenamefont {Fontana}, \citenamefont {Cuttle}, \citenamefont {Pihler-Puzovi{\'c}}, \citenamefont {Hazel},\ and\ \citenamefont {Juel}}]{fontana2024peeling}%
  \BibitemOpen
  \bibfield  {author} {\bibinfo {author} {\bibfnamefont {J.~V.}\ \bibnamefont {Fontana}}, \bibinfo {author} {\bibfnamefont {C.}~\bibnamefont {Cuttle}}, \bibinfo {author} {\bibfnamefont {D.}~\bibnamefont {Pihler-Puzovi{\'c}}}, \bibinfo {author} {\bibfnamefont {A.~L.}\ \bibnamefont {Hazel}},\ and\ \bibinfo {author} {\bibfnamefont {A.}~\bibnamefont {Juel}},\ }\bibfield  {title} {\bibinfo {title} {{Peeling fingers in an elastic Hele-Shaw channel}},\ }\href@noop {} {\bibfield  {journal} {\bibinfo  {journal} {J. Fluid Mech.}\ }\textbf {\bibinfo {volume} {985}},\ \bibinfo {pages} {A1} (\bibinfo {year} {2024})}\BibitemShut {NoStop}%
\bibitem [{\citenamefont {Gervais}\ \emph {et~al.}(2006)\citenamefont {Gervais}, \citenamefont {El-Ali}, \citenamefont {G{\"u}nther},\ and\ \citenamefont {Jensen}}]{gervais2006flow}%
  \BibitemOpen
  \bibfield  {author} {\bibinfo {author} {\bibfnamefont {T.}~\bibnamefont {Gervais}}, \bibinfo {author} {\bibfnamefont {J.}~\bibnamefont {El-Ali}}, \bibinfo {author} {\bibfnamefont {A.}~\bibnamefont {G{\"u}nther}},\ and\ \bibinfo {author} {\bibfnamefont {K.~F.}\ \bibnamefont {Jensen}},\ }\bibfield  {title} {\bibinfo {title} {Flow-induced deformation of shallow microfluidic channels},\ }\href@noop {} {\bibfield  {journal} {\bibinfo  {journal} {Lab Chip}\ }\textbf {\bibinfo {volume} {6}},\ \bibinfo {pages} {500} (\bibinfo {year} {2006})}\BibitemShut {NoStop}%
\bibitem [{\citenamefont {Dendukuri}\ \emph {et~al.}(2007)\citenamefont {Dendukuri}, \citenamefont {Gu}, \citenamefont {Pregibon}, \citenamefont {Hatton},\ and\ \citenamefont {Doyle}}]{dendukuri2007stop}%
  \BibitemOpen
  \bibfield  {author} {\bibinfo {author} {\bibfnamefont {D.}~\bibnamefont {Dendukuri}}, \bibinfo {author} {\bibfnamefont {S.~S.}\ \bibnamefont {Gu}}, \bibinfo {author} {\bibfnamefont {D.~C.}\ \bibnamefont {Pregibon}}, \bibinfo {author} {\bibfnamefont {T.~A.}\ \bibnamefont {Hatton}},\ and\ \bibinfo {author} {\bibfnamefont {P.~S.}\ \bibnamefont {Doyle}},\ }\bibfield  {title} {\bibinfo {title} {Stop-flow lithography in a microfluidic device},\ }\href@noop {} {\bibfield  {journal} {\bibinfo  {journal} {Lab Chip}\ }\textbf {\bibinfo {volume} {7}},\ \bibinfo {pages} {818} (\bibinfo {year} {2007})}\BibitemShut {NoStop}%
\bibitem [{\citenamefont {Hardy}\ \emph {et~al.}(2009)\citenamefont {Hardy}, \citenamefont {Uechi}, \citenamefont {Zhen},\ and\ \citenamefont {Kavehpour}}]{hardy2009deformation}%
  \BibitemOpen
  \bibfield  {author} {\bibinfo {author} {\bibfnamefont {B.~S.}\ \bibnamefont {Hardy}}, \bibinfo {author} {\bibfnamefont {K.}~\bibnamefont {Uechi}}, \bibinfo {author} {\bibfnamefont {J.}~\bibnamefont {Zhen}},\ and\ \bibinfo {author} {\bibfnamefont {H.~P.}\ \bibnamefont {Kavehpour}},\ }\bibfield  {title} {\bibinfo {title} {The deformation of flexible {PDMS} microchannels under a pressure driven flow},\ }\href@noop {} {\bibfield  {journal} {\bibinfo  {journal} {Lab Chip}\ }\textbf {\bibinfo {volume} {9}},\ \bibinfo {pages} {935} (\bibinfo {year} {2009})}\BibitemShut {NoStop}%
\bibitem [{\citenamefont {Christov}\ \emph {et~al.}(2018)\citenamefont {Christov}, \citenamefont {Cognet}, \citenamefont {Shidhore},\ and\ \citenamefont {Stone}}]{christov2018flow}%
  \BibitemOpen
  \bibfield  {author} {\bibinfo {author} {\bibfnamefont {I.~C.}\ \bibnamefont {Christov}}, \bibinfo {author} {\bibfnamefont {V.}~\bibnamefont {Cognet}}, \bibinfo {author} {\bibfnamefont {T.~C.}\ \bibnamefont {Shidhore}},\ and\ \bibinfo {author} {\bibfnamefont {H.~A.}\ \bibnamefont {Stone}},\ }\bibfield  {title} {\bibinfo {title} {Flow rate--pressure drop relation for deformable shallow microfluidic channels},\ }\href@noop {} {\bibfield  {journal} {\bibinfo  {journal} {J. Fluid Mech.}\ }\textbf {\bibinfo {volume} {841}},\ \bibinfo {pages} {267} (\bibinfo {year} {2018})}\BibitemShut {NoStop}%
\bibitem [{\citenamefont {Shepherd}\ \emph {et~al.}(2011)\citenamefont {Shepherd}, \citenamefont {Ilievski}, \citenamefont {Choi}, \citenamefont {Morin}, \citenamefont {Stokes}, \citenamefont {Mazzeo}, \citenamefont {Chen}, \citenamefont {Wang},\ and\ \citenamefont {Whitesides}}]{shepherd2011multigait}%
  \BibitemOpen
  \bibfield  {author} {\bibinfo {author} {\bibfnamefont {R.~F.}\ \bibnamefont {Shepherd}}, \bibinfo {author} {\bibfnamefont {F.}~\bibnamefont {Ilievski}}, \bibinfo {author} {\bibfnamefont {W.}~\bibnamefont {Choi}}, \bibinfo {author} {\bibfnamefont {S.~A.}\ \bibnamefont {Morin}}, \bibinfo {author} {\bibfnamefont {A.~A.}\ \bibnamefont {Stokes}}, \bibinfo {author} {\bibfnamefont {A.~D.}\ \bibnamefont {Mazzeo}}, \bibinfo {author} {\bibfnamefont {X.}~\bibnamefont {Chen}}, \bibinfo {author} {\bibfnamefont {M.}~\bibnamefont {Wang}},\ and\ \bibinfo {author} {\bibfnamefont {G.~M.}\ \bibnamefont {Whitesides}},\ }\bibfield  {title} {\bibinfo {title} {Multigait soft robot},\ }\href@noop {} {\bibfield  {journal} {\bibinfo  {journal} {Proc. Natl Acad. of Sci. USA}\ }\textbf {\bibinfo {volume} {108}},\ \bibinfo {pages} {20400} (\bibinfo {year} {2011})}\BibitemShut {NoStop}%
\bibitem [{\citenamefont {Elbaz}\ and\ \citenamefont {Gat}(2014)}]{elbaz2014dynamics}%
  \BibitemOpen
  \bibfield  {author} {\bibinfo {author} {\bibfnamefont {S.~B.}\ \bibnamefont {Elbaz}}\ and\ \bibinfo {author} {\bibfnamefont {A.~D.}\ \bibnamefont {Gat}},\ }\bibfield  {title} {\bibinfo {title} {Dynamics of viscous liquid within a closed elastic cylinder subject to external forces with application to soft robotics},\ }\href@noop {} {\bibfield  {journal} {\bibinfo  {journal} {J. Fluid Mech.}\ }\textbf {\bibinfo {volume} {758}},\ \bibinfo {pages} {221} (\bibinfo {year} {2014})}\BibitemShut {NoStop}%
\bibitem [{\citenamefont {Rus}\ and\ \citenamefont {Tolley}(2015)}]{rus2015design}%
  \BibitemOpen
  \bibfield  {author} {\bibinfo {author} {\bibfnamefont {D.}~\bibnamefont {Rus}}\ and\ \bibinfo {author} {\bibfnamefont {M.~T.}\ \bibnamefont {Tolley}},\ }\bibfield  {title} {\bibinfo {title} {Design, fabrication and control of soft robots},\ }\href@noop {} {\bibfield  {journal} {\bibinfo  {journal} {Nature}\ }\textbf {\bibinfo {volume} {521}},\ \bibinfo {pages} {467} (\bibinfo {year} {2015})}\BibitemShut {NoStop}%
\bibitem [{\citenamefont {Polygerinos}\ \emph {et~al.}(2017)\citenamefont {Polygerinos}, \citenamefont {Correll}, \citenamefont {Morin}, \citenamefont {Mosadegh}, \citenamefont {Onal}, \citenamefont {Petersen}, \citenamefont {Cianchetti}, \citenamefont {Tolley},\ and\ \citenamefont {Shepherd}}]{polygerinos2017soft}%
  \BibitemOpen
  \bibfield  {author} {\bibinfo {author} {\bibfnamefont {P.}~\bibnamefont {Polygerinos}}, \bibinfo {author} {\bibfnamefont {N.}~\bibnamefont {Correll}}, \bibinfo {author} {\bibfnamefont {S.~A.}\ \bibnamefont {Morin}}, \bibinfo {author} {\bibfnamefont {B.}~\bibnamefont {Mosadegh}}, \bibinfo {author} {\bibfnamefont {C.~D.}\ \bibnamefont {Onal}}, \bibinfo {author} {\bibfnamefont {K.}~\bibnamefont {Petersen}}, \bibinfo {author} {\bibfnamefont {M.}~\bibnamefont {Cianchetti}}, \bibinfo {author} {\bibfnamefont {M.~T.}\ \bibnamefont {Tolley}},\ and\ \bibinfo {author} {\bibfnamefont {R.~F.}\ \bibnamefont {Shepherd}},\ }\bibfield  {title} {\bibinfo {title} {{Soft robotics: Review of fluid-driven intrinsically soft devices; manufacturing, sensing, control, and applications in human-robot interaction}},\ }\href@noop {} {\bibfield  {journal} {\bibinfo  {journal} {Adv. Eng. Mater.}\ }\textbf {\bibinfo {volume} {19}},\ \bibinfo {pages} {1700016} (\bibinfo {year} {2017})}\BibitemShut {NoStop}%
\bibitem [{\citenamefont {Matia}\ \emph {et~al.}(2017)\citenamefont {Matia}, \citenamefont {Elimelech},\ and\ \citenamefont {Gat}}]{MEG17}%
  \BibitemOpen
  \bibfield  {author} {\bibinfo {author} {\bibfnamefont {Y.}~\bibnamefont {Matia}}, \bibinfo {author} {\bibfnamefont {T.}~\bibnamefont {Elimelech}},\ and\ \bibinfo {author} {\bibfnamefont {A.~D.}\ \bibnamefont {Gat}},\ }\bibfield  {title} {\bibinfo {title} {{Leveraging internal viscous flow to extend the capabilities of beam-shaped soft robotic actuators}},\ }\href@noop {} {\bibfield  {journal} {\bibinfo  {journal} {Soft Robotics}\ }\textbf {\bibinfo {volume} {4}},\ \bibinfo {pages} {126} (\bibinfo {year} {2017})}\BibitemShut {NoStop}%
\bibitem [{\citenamefont {Salem}\ \emph {et~al.}(2020)\citenamefont {Salem}, \citenamefont {Gamus}, \citenamefont {Or},\ and\ \citenamefont {Gat}}]{salem2020leveraging}%
  \BibitemOpen
  \bibfield  {author} {\bibinfo {author} {\bibfnamefont {L.}~\bibnamefont {Salem}}, \bibinfo {author} {\bibfnamefont {B.}~\bibnamefont {Gamus}}, \bibinfo {author} {\bibfnamefont {Y.}~\bibnamefont {Or}},\ and\ \bibinfo {author} {\bibfnamefont {A.~D.}\ \bibnamefont {Gat}},\ }\bibfield  {title} {\bibinfo {title} {Leveraging viscous peeling to create and activate soft actuators and microfluidic devices},\ }\href@noop {} {\bibfield  {journal} {\bibinfo  {journal} {Soft Robotics}\ }\textbf {\bibinfo {volume} {7}},\ \bibinfo {pages} {76} (\bibinfo {year} {2020})}\BibitemShut {NoStop}%
\bibitem [{\citenamefont {Park}\ \emph {et~al.}(2021)\citenamefont {Park}, \citenamefont {Tixier}, \citenamefont {Paludan}, \citenamefont {{\O}stergaard}, \citenamefont {Zwieniecki},\ and\ \citenamefont {Jensen}}]{park2021fluid}%
  \BibitemOpen
  \bibfield  {author} {\bibinfo {author} {\bibfnamefont {K.}~\bibnamefont {Park}}, \bibinfo {author} {\bibfnamefont {A.}~\bibnamefont {Tixier}}, \bibinfo {author} {\bibfnamefont {M.}~\bibnamefont {Paludan}}, \bibinfo {author} {\bibfnamefont {E.}~\bibnamefont {{\O}stergaard}}, \bibinfo {author} {\bibfnamefont {M.}~\bibnamefont {Zwieniecki}},\ and\ \bibinfo {author} {\bibfnamefont {K.~H.}\ \bibnamefont {Jensen}},\ }\bibfield  {title} {\bibinfo {title} {Fluid-structure interactions enable passive flow control in real and biomimetic plants},\ }\href@noop {} {\bibfield  {journal} {\bibinfo  {journal} {Phys. Rev. Fluids}\ }\textbf {\bibinfo {volume} {6}},\ \bibinfo {pages} {123102} (\bibinfo {year} {2021})}\BibitemShut {NoStop}%
\bibitem [{\citenamefont {Boyko}\ \emph {et~al.}(2022)\citenamefont {Boyko}, \citenamefont {Stone},\ and\ \citenamefont {Christov}}]{boyko2022flow}%
  \BibitemOpen
  \bibfield  {author} {\bibinfo {author} {\bibfnamefont {E.}~\bibnamefont {Boyko}}, \bibinfo {author} {\bibfnamefont {H.~A.}\ \bibnamefont {Stone}},\ and\ \bibinfo {author} {\bibfnamefont {I.~C.}\ \bibnamefont {Christov}},\ }\bibfield  {title} {\bibinfo {title} {Flow rate-pressure drop relation for deformable channels via fluidic and elastic reciprocal theorems},\ }\href@noop {} {\bibfield  {journal} {\bibinfo  {journal} {Phys. Rev. Fluids}\ }\textbf {\bibinfo {volume} {7}},\ \bibinfo {pages} {L092201} (\bibinfo {year} {2022})}\BibitemShut {NoStop}%
\bibitem [{\citenamefont {Mart{\'\i}nez-Calvo}\ \emph {et~al.}(2020)\citenamefont {Mart{\'\i}nez-Calvo}, \citenamefont {Sevilla}, \citenamefont {Peng},\ and\ \citenamefont {Stone}}]{martinez2020start}%
  \BibitemOpen
  \bibfield  {author} {\bibinfo {author} {\bibfnamefont {A.}~\bibnamefont {Mart{\'\i}nez-Calvo}}, \bibinfo {author} {\bibfnamefont {A.}~\bibnamefont {Sevilla}}, \bibinfo {author} {\bibfnamefont {G.~G.}\ \bibnamefont {Peng}},\ and\ \bibinfo {author} {\bibfnamefont {H.~A.}\ \bibnamefont {Stone}},\ }\bibfield  {title} {\bibinfo {title} {Start-up flow in shallow deformable microchannels},\ }\href@noop {} {\bibfield  {journal} {\bibinfo  {journal} {J. Fluid Mech.}\ }\textbf {\bibinfo {volume} {885}},\ \bibinfo {pages} {A25} (\bibinfo {year} {2020})}\BibitemShut {NoStop}%
\bibitem [{\citenamefont {Guyard}\ \emph {et~al.}(2022)\citenamefont {Guyard}, \citenamefont {Restagno},\ and\ \citenamefont {McGraw}}]{guyard2022elastohydrodynamic}%
  \BibitemOpen
  \bibfield  {author} {\bibinfo {author} {\bibfnamefont {G.}~\bibnamefont {Guyard}}, \bibinfo {author} {\bibfnamefont {F.}~\bibnamefont {Restagno}},\ and\ \bibinfo {author} {\bibfnamefont {J.~D.}\ \bibnamefont {McGraw}},\ }\bibfield  {title} {\bibinfo {title} {Elastohydrodynamic relaxation of soft and deformable microchannels},\ }\href@noop {} {\bibfield  {journal} {\bibinfo  {journal} {Phys. Rev. Lett.}\ }\textbf {\bibinfo {volume} {129}},\ \bibinfo {pages} {204501} (\bibinfo {year} {2022})}\BibitemShut {NoStop}%
\bibitem [{\citenamefont {Biviano}\ \emph {et~al.}(2022)\citenamefont {Biviano}, \citenamefont {Paludan}, \citenamefont {Christensen}, \citenamefont {{\O}stergaard},\ and\ \citenamefont {Jensen}}]{biviano2022smoothing}%
  \BibitemOpen
  \bibfield  {author} {\bibinfo {author} {\bibfnamefont {M.~D.}\ \bibnamefont {Biviano}}, \bibinfo {author} {\bibfnamefont {M.~V.}\ \bibnamefont {Paludan}}, \bibinfo {author} {\bibfnamefont {A.~H.}\ \bibnamefont {Christensen}}, \bibinfo {author} {\bibfnamefont {E.~V.}\ \bibnamefont {{\O}stergaard}},\ and\ \bibinfo {author} {\bibfnamefont {K.~H.}\ \bibnamefont {Jensen}},\ }\bibfield  {title} {\bibinfo {title} {Smoothing oscillatory peristaltic pump flow with bioinspired passive components},\ }\href@noop {} {\bibfield  {journal} {\bibinfo  {journal} {Phys. Rev. Appl.}\ }\textbf {\bibinfo {volume} {18}},\ \bibinfo {pages} {064013} (\bibinfo {year} {2022})}\BibitemShut {NoStop}%
\bibitem [{\citenamefont {Mukherjee}\ \emph {et~al.}(2013)\citenamefont {Mukherjee}, \citenamefont {Chakraborty},\ and\ \citenamefont {Chakraborty}}]{mukherjee2013relaxation}%
  \BibitemOpen
  \bibfield  {author} {\bibinfo {author} {\bibfnamefont {U.}~\bibnamefont {Mukherjee}}, \bibinfo {author} {\bibfnamefont {J.}~\bibnamefont {Chakraborty}},\ and\ \bibinfo {author} {\bibfnamefont {S.}~\bibnamefont {Chakraborty}},\ }\bibfield  {title} {\bibinfo {title} {Relaxation characteristics of a compliant microfluidic channel under electroosmotic flow},\ }\href@noop {} {\bibfield  {journal} {\bibinfo  {journal} {Soft Matt.}\ }\textbf {\bibinfo {volume} {9}},\ \bibinfo {pages} {1562} (\bibinfo {year} {2013})}\BibitemShut {NoStop}%
\bibitem [{\citenamefont {de~Rutte}\ \emph {et~al.}(2016)\citenamefont {de~Rutte}, \citenamefont {Janssen}, \citenamefont {Tas}, \citenamefont {Eijkel},\ and\ \citenamefont {Pennathur}}]{de2016numerical}%
  \BibitemOpen
  \bibfield  {author} {\bibinfo {author} {\bibfnamefont {J.~M.}\ \bibnamefont {de~Rutte}}, \bibinfo {author} {\bibfnamefont {K.~G.~H.}\ \bibnamefont {Janssen}}, \bibinfo {author} {\bibfnamefont {N.~R.}\ \bibnamefont {Tas}}, \bibinfo {author} {\bibfnamefont {J.~C.~T.}\ \bibnamefont {Eijkel}},\ and\ \bibinfo {author} {\bibfnamefont {S.}~\bibnamefont {Pennathur}},\ }\bibfield  {title} {\bibinfo {title} {Numerical investigation of micro-and nanochannel deformation due to discontinuous electroosmotic flow},\ }\href@noop {} {\bibfield  {journal} {\bibinfo  {journal} {Microfluid. Nanofluid.}\ }\textbf {\bibinfo {volume} {20}},\ \bibinfo {pages} {150} (\bibinfo {year} {2016})}\BibitemShut {NoStop}%
\bibitem [{\citenamefont {Rubin}\ \emph {et~al.}(2017)\citenamefont {Rubin}, \citenamefont {Tulchinsky}, \citenamefont {Gat},\ and\ \citenamefont {Bercovici}}]{rubin2017elastic}%
  \BibitemOpen
  \bibfield  {author} {\bibinfo {author} {\bibfnamefont {S.}~\bibnamefont {Rubin}}, \bibinfo {author} {\bibfnamefont {A.}~\bibnamefont {Tulchinsky}}, \bibinfo {author} {\bibfnamefont {A.~D.}\ \bibnamefont {Gat}},\ and\ \bibinfo {author} {\bibfnamefont {M.}~\bibnamefont {Bercovici}},\ }\bibfield  {title} {\bibinfo {title} {Elastic deformations driven by non-uniform lubrication flows},\ }\href@noop {} {\bibfield  {journal} {\bibinfo  {journal} {J. Fluid Mech.}\ }\textbf {\bibinfo {volume} {812}},\ \bibinfo {pages} {841} (\bibinfo {year} {2017})}\BibitemShut {NoStop}%
\bibitem [{\citenamefont {Boyko}\ \emph {et~al.}(2019)\citenamefont {Boyko}, \citenamefont {Eshel}, \citenamefont {Gommed}, \citenamefont {Gat},\ and\ \citenamefont {Bercovici}}]{boyko2019elastohydrodynamics}%
  \BibitemOpen
  \bibfield  {author} {\bibinfo {author} {\bibfnamefont {E.}~\bibnamefont {Boyko}}, \bibinfo {author} {\bibfnamefont {R.}~\bibnamefont {Eshel}}, \bibinfo {author} {\bibfnamefont {K.}~\bibnamefont {Gommed}}, \bibinfo {author} {\bibfnamefont {A.~D.}\ \bibnamefont {Gat}},\ and\ \bibinfo {author} {\bibfnamefont {M.}~\bibnamefont {Bercovici}},\ }\bibfield  {title} {\bibinfo {title} {Elastohydrodynamics of a pre-stretched finite elastic sheet lubricated by a thin viscous film with application to microfluidic soft actuators},\ }\href@noop {} {\bibfield  {journal} {\bibinfo  {journal} {J. Fluid Mech.}\ }\textbf {\bibinfo {volume} {862}},\ \bibinfo {pages} {732} (\bibinfo {year} {2019})}\BibitemShut {NoStop}%
\bibitem [{\citenamefont {Rallabandi}(2024)}]{rallabandi2024fluid}%
  \BibitemOpen
  \bibfield  {author} {\bibinfo {author} {\bibfnamefont {B.}~\bibnamefont {Rallabandi}},\ }\bibfield  {title} {\bibinfo {title} {{Fluid-elastic interactions near contact at low Reynolds number}},\ }\href@noop {} {\bibfield  {journal} {\bibinfo  {journal} {Annu. Rev. Fluid Mech.}\ }\textbf {\bibinfo {volume} {56}},\ \bibinfo {pages} {491} (\bibinfo {year} {2024})}\BibitemShut {NoStop}%
\bibitem [{\citenamefont {Heil}\ and\ \citenamefont {Hazel}(2011)}]{heil2011fluid}%
  \BibitemOpen
  \bibfield  {author} {\bibinfo {author} {\bibfnamefont {M.}~\bibnamefont {Heil}}\ and\ \bibinfo {author} {\bibfnamefont {A.~L.}\ \bibnamefont {Hazel}},\ }\bibfield  {title} {\bibinfo {title} {Fluid-structure interaction in internal physiological flows},\ }\href@noop {} {\bibfield  {journal} {\bibinfo  {journal} {Annu. Rev. Fluid Mech.}\ }\textbf {\bibinfo {volume} {43}},\ \bibinfo {pages} {141} (\bibinfo {year} {2011})}\BibitemShut {NoStop}%
\bibitem [{\citenamefont {Juel}\ \emph {et~al.}(2018)\citenamefont {Juel}, \citenamefont {Pihler-Puzovi{\'c}},\ and\ \citenamefont {Heil}}]{juel2018instabilities}%
  \BibitemOpen
  \bibfield  {author} {\bibinfo {author} {\bibfnamefont {A.}~\bibnamefont {Juel}}, \bibinfo {author} {\bibfnamefont {D.}~\bibnamefont {Pihler-Puzovi{\'c}}},\ and\ \bibinfo {author} {\bibfnamefont {M.}~\bibnamefont {Heil}},\ }\bibfield  {title} {\bibinfo {title} {Instabilities in blistering},\ }\href@noop {} {\bibfield  {journal} {\bibinfo  {journal} {Annu. Rev. Fluid Mech.}\ }\textbf {\bibinfo {volume} {50}},\ \bibinfo {pages} {691} (\bibinfo {year} {2018})}\BibitemShut {NoStop}%
\bibitem [{\citenamefont {Heil}(1997)}]{heil1997stokes}%
  \BibitemOpen
  \bibfield  {author} {\bibinfo {author} {\bibfnamefont {M.}~\bibnamefont {Heil}},\ }\bibfield  {title} {\bibinfo {title} {Stokes flow in collapsible tubes: computation and experiment},\ }\href@noop {} {\bibfield  {journal} {\bibinfo  {journal} {J. Fluid Mech.}\ }\textbf {\bibinfo {volume} {353}},\ \bibinfo {pages} {285} (\bibinfo {year} {1997})}\BibitemShut {NoStop}%
\bibitem [{\citenamefont {Gomez}\ \emph {et~al.}(2017)\citenamefont {Gomez}, \citenamefont {Moulton},\ and\ \citenamefont {Vella}}]{gomez2017passive}%
  \BibitemOpen
  \bibfield  {author} {\bibinfo {author} {\bibfnamefont {M.}~\bibnamefont {Gomez}}, \bibinfo {author} {\bibfnamefont {D.~E.}\ \bibnamefont {Moulton}},\ and\ \bibinfo {author} {\bibfnamefont {D.}~\bibnamefont {Vella}},\ }\bibfield  {title} {\bibinfo {title} {Passive control of viscous flow via elastic snap-through},\ }\href@noop {} {\bibfield  {journal} {\bibinfo  {journal} {Phys. Rev. Lett.}\ }\textbf {\bibinfo {volume} {119}},\ \bibinfo {pages} {144502} (\bibinfo {year} {2017})}\BibitemShut {NoStop}%
\bibitem [{\citenamefont {Kodio}\ \emph {et~al.}(2017)\citenamefont {Kodio}, \citenamefont {Griffiths},\ and\ \citenamefont {Vella}}]{kodio2017lubricated}%
  \BibitemOpen
  \bibfield  {author} {\bibinfo {author} {\bibfnamefont {O.}~\bibnamefont {Kodio}}, \bibinfo {author} {\bibfnamefont {I.~M.}\ \bibnamefont {Griffiths}},\ and\ \bibinfo {author} {\bibfnamefont {D.}~\bibnamefont {Vella}},\ }\bibfield  {title} {\bibinfo {title} {{Lubricated wrinkles: Imposed constraints affect the dynamics of wrinkle coarsening}},\ }\href@noop {} {\bibfield  {journal} {\bibinfo  {journal} {Phys. Rev. Fluids}\ }\textbf {\bibinfo {volume} {2}},\ \bibinfo {pages} {014202} (\bibinfo {year} {2017})}\BibitemShut {NoStop}%
\bibitem [{\citenamefont {Peretz}\ \emph {et~al.}(2020)\citenamefont {Peretz}, \citenamefont {Mishra}, \citenamefont {Shepherd},\ and\ \citenamefont {Gat}}]{peretz2020underactuated}%
  \BibitemOpen
  \bibfield  {author} {\bibinfo {author} {\bibfnamefont {O.}~\bibnamefont {Peretz}}, \bibinfo {author} {\bibfnamefont {A.~K.}\ \bibnamefont {Mishra}}, \bibinfo {author} {\bibfnamefont {R.~F.}\ \bibnamefont {Shepherd}},\ and\ \bibinfo {author} {\bibfnamefont {A.~D.}\ \bibnamefont {Gat}},\ }\bibfield  {title} {\bibinfo {title} {Underactuated fluidic control of a continuous multistable membrane},\ }\href@noop {} {\bibfield  {journal} {\bibinfo  {journal} {Proc. Natl Acad. Sci. USA}\ }\textbf {\bibinfo {volume} {117}},\ \bibinfo {pages} {5217} (\bibinfo {year} {2020})}\BibitemShut {NoStop}%
\bibitem [{\citenamefont {Box}\ \emph {et~al.}(2020)\citenamefont {Box}, \citenamefont {Peng}, \citenamefont {Pihler-Puzovi{\'c}},\ and\ \citenamefont {Juel}}]{box2020flow}%
  \BibitemOpen
  \bibfield  {author} {\bibinfo {author} {\bibfnamefont {F.}~\bibnamefont {Box}}, \bibinfo {author} {\bibfnamefont {G.~G.}\ \bibnamefont {Peng}}, \bibinfo {author} {\bibfnamefont {D.}~\bibnamefont {Pihler-Puzovi{\'c}}},\ and\ \bibinfo {author} {\bibfnamefont {A.}~\bibnamefont {Juel}},\ }\bibfield  {title} {\bibinfo {title} {{Flow-induced choking of a compliant Hele-Shaw cell}},\ }\href@noop {} {\bibfield  {journal} {\bibinfo  {journal} {Proc. Natl Acad. Sci. USA}\ }\textbf {\bibinfo {volume} {117}},\ \bibinfo {pages} {30228} (\bibinfo {year} {2020})}\BibitemShut {NoStop}%
\bibitem [{\citenamefont {Boyko}\ \emph {et~al.}(2020{\natexlab{a}})\citenamefont {Boyko}, \citenamefont {Eshel}, \citenamefont {Gat},\ and\ \citenamefont {Bercovici}}]{boyko2020nonuniform}%
  \BibitemOpen
  \bibfield  {author} {\bibinfo {author} {\bibfnamefont {E.}~\bibnamefont {Boyko}}, \bibinfo {author} {\bibfnamefont {R.}~\bibnamefont {Eshel}}, \bibinfo {author} {\bibfnamefont {A.~D.}\ \bibnamefont {Gat}},\ and\ \bibinfo {author} {\bibfnamefont {M.}~\bibnamefont {Bercovici}},\ }\bibfield  {title} {\bibinfo {title} {Nonuniform electro-osmotic flow drives fluid-structure instability},\ }\href@noop {} {\bibfield  {journal} {\bibinfo  {journal} {Phys. Rev. Lett.}\ }\textbf {\bibinfo {volume} {124}},\ \bibinfo {pages} {024501} (\bibinfo {year} {2020}{\natexlab{a}})}\BibitemShut {NoStop}%
\bibitem [{\citenamefont {Boyko}\ \emph {et~al.}(2020{\natexlab{b}})\citenamefont {Boyko}, \citenamefont {Ilssar}, \citenamefont {Bercovici},\ and\ \citenamefont {Gat}}]{boyko2020interfacial}%
  \BibitemOpen
  \bibfield  {author} {\bibinfo {author} {\bibfnamefont {E.}~\bibnamefont {Boyko}}, \bibinfo {author} {\bibfnamefont {D.}~\bibnamefont {Ilssar}}, \bibinfo {author} {\bibfnamefont {M.}~\bibnamefont {Bercovici}},\ and\ \bibinfo {author} {\bibfnamefont {A.~D.}\ \bibnamefont {Gat}},\ }\bibfield  {title} {\bibinfo {title} {Interfacial instability of thin films in soft microfluidic configurations actuated by electro-osmotic flow},\ }\href@noop {} {\bibfield  {journal} {\bibinfo  {journal} {Phys. Rev. Fluids}\ }\textbf {\bibinfo {volume} {5}},\ \bibinfo {pages} {104201} (\bibinfo {year} {2020}{\natexlab{b}})}\BibitemShut {NoStop}%
\bibitem [{\citenamefont {Leal}(2007)}]{leal2007advanced}%
  \BibitemOpen
  \bibfield  {author} {\bibinfo {author} {\bibfnamefont {L.~G.}\ \bibnamefont {Leal}},\ }\href@noop {} {\emph {\bibinfo {title} {Advanced {T}ransport {P}henomena: {F}luid {M}echanics and {C}onvective {T}ransport {P}processes}}}\ (\bibinfo  {publisher} {Cambridge University Press},\ \bibinfo {year} {2007})\BibitemShut {NoStop}%
\bibitem [{\citenamefont {Migacz}\ and\ \citenamefont {Ault}(2022)}]{migacz2022diffusiophoresis}%
  \BibitemOpen
  \bibfield  {author} {\bibinfo {author} {\bibfnamefont {R.~E.}\ \bibnamefont {Migacz}}\ and\ \bibinfo {author} {\bibfnamefont {J.~T.}\ \bibnamefont {Ault}},\ }\bibfield  {title} {\bibinfo {title} {{Diffusiophoresis in a Taylor-dispersing solute}},\ }\href@noop {} {\bibfield  {journal} {\bibinfo  {journal} {Phys. Rev. Fluids}\ }\textbf {\bibinfo {volume} {7}},\ \bibinfo {pages} {034202} (\bibinfo {year} {2022})}\BibitemShut {NoStop}%
\bibitem [{\citenamefont {Lee}\ \emph {et~al.}(2023)\citenamefont {Lee}, \citenamefont {Lee},\ and\ \citenamefont {Ault}}]{lee2023role}%
  \BibitemOpen
  \bibfield  {author} {\bibinfo {author} {\bibfnamefont {S.}~\bibnamefont {Lee}}, \bibinfo {author} {\bibfnamefont {J.}~\bibnamefont {Lee}},\ and\ \bibinfo {author} {\bibfnamefont {J.~T.}\ \bibnamefont {Ault}},\ }\bibfield  {title} {\bibinfo {title} {The role of variable zeta potential on diffusiophoretic and diffusioosmotic transport},\ }\href@noop {} {\bibfield  {journal} {\bibinfo  {journal} {Colloids Surf. A}\ }\textbf {\bibinfo {volume} {659}},\ \bibinfo {pages} {130775} (\bibinfo {year} {2023})}\BibitemShut {NoStop}%
\bibitem [{\citenamefont {Rebeiz}(2004)}]{rebeiz2004rf}%
  \BibitemOpen
  \bibfield  {author} {\bibinfo {author} {\bibfnamefont {G.~M.}\ \bibnamefont {Rebeiz}},\ }\href@noop {} {\emph {\bibinfo {title} {RF MEMS: Theory, Design, and Technology}}}\ (\bibinfo  {publisher} {John Wiley \& Sons},\ \bibinfo {year} {2004})\BibitemShut {NoStop}%
\bibitem [{\citenamefont {Singh}\ \emph {et~al.}(2014)\citenamefont {Singh}, \citenamefont {Lister},\ and\ \citenamefont {Vella}}]{singh2014fluid}%
  \BibitemOpen
  \bibfield  {author} {\bibinfo {author} {\bibfnamefont {K.}~\bibnamefont {Singh}}, \bibinfo {author} {\bibfnamefont {J.~R.}\ \bibnamefont {Lister}},\ and\ \bibinfo {author} {\bibfnamefont {D.}~\bibnamefont {Vella}},\ }\bibfield  {title} {\bibinfo {title} {A fluid-mechanical model of elastocapillary coalescence},\ }\href@noop {} {\bibfield  {journal} {\bibinfo  {journal} {J. Fluid Mech.}\ }\textbf {\bibinfo {volume} {745}},\ \bibinfo {pages} {621} (\bibinfo {year} {2014})}\BibitemShut {NoStop}%
\bibitem [{\citenamefont {Van{\`y}sek}(1993)}]{vanysek1993ionic}%
  \BibitemOpen
  \bibfield  {author} {\bibinfo {author} {\bibfnamefont {P.}~\bibnamefont {Van{\`y}sek}},\ }\bibfield  {title} {\bibinfo {title} {Ionic conductivity and diffusion at infinite dilution},\ }in\ \href@noop {} {\emph {\bibinfo {booktitle} {CRC Handbook of Chemistry and Physics}}}\ (\bibinfo  {publisher} {CRC Press},\ \bibinfo {year} {1993})\ pp.\ \bibinfo {pages} {5--92}\BibitemShut {NoStop}%
\bibitem [{\citenamefont {Gomez}\ \emph {et~al.}(2018)\citenamefont {Gomez}, \citenamefont {Moulton},\ and\ \citenamefont {Vella}}]{gomez2018delayed}%
  \BibitemOpen
  \bibfield  {author} {\bibinfo {author} {\bibfnamefont {M.}~\bibnamefont {Gomez}}, \bibinfo {author} {\bibfnamefont {D.~E.}\ \bibnamefont {Moulton}},\ and\ \bibinfo {author} {\bibfnamefont {D.}~\bibnamefont {Vella}},\ }\bibfield  {title} {\bibinfo {title} {{Delayed pull-in transitions in overdamped MEMS devices}},\ }\href@noop {} {\bibfield  {journal} {\bibinfo  {journal} {J. Micromech. Microeng.}\ }\textbf {\bibinfo {volume} {28}},\ \bibinfo {pages} {015006} (\bibinfo {year} {2018})}\BibitemShut {NoStop}%
\end{thebibliography}%

\end{document}